\documentclass[12pt]{iopart}
\usepackage{iopams}
\expandafter\let\csname equation*\endcsname\relax
\expandafter\let\csname endequation*\endcsname\relax
\usepackage{amsmath}

\usepackage[numbers,sort&compress]{natbib}

\usepackage{amsfonts}
\usepackage{amssymb}
\usepackage{amsthm}
\usepackage{graphicx}
\usepackage{verbatim}
\usepackage{hyperref}
\usepackage[english]{babel}
\usepackage[pdftex]{color}
\usepackage[dvipsnames]{xcolor}

\hypersetup{
    colorlinks=true,       
    linkcolor=OliveGreen,          
    citecolor=Cerulean,        
    filecolor=BrickRed,      
    urlcolor=RoyalBlue,           
    runcolor=Cerulean
}

\graphicspath{{./imgs/}}

\newcommand{\bra}[1]{\langle#1|}
\newcommand{\ket}[1]{|#1\rangle}

\newcommand{\ketbra}[2]{|#1\rangle \langle #2 |}

\usepackage{quantikz}
\usepackage{physics}
\usepackage{soul}

\makeatletter
\newcommand\thefontsize{The current font size is: \f@size pt}
\makeatother

\def\correspondingauthor{\footnote{Corresponding author.}}

\usepackage{subcaption}

\begin{document}

\title{Benchmarking the performance of a high-Q cavity qudit using random unitaries}

\author{Nicholas Bornman\correspondingauthor}
\address{Superconducting Quantum Materials and Systems Center, Fermi National Accelerator Laboratory, Batavia, IL 60510, USA}
\ead{nbornman@fnal.gov}
\author{Tanay Roy}
\ead{roytanay@fnal.gov}
\address{Superconducting Quantum Materials and Systems Center, Fermi National Accelerator Laboratory, Batavia, IL 60510, USA}
\author{Joshua A Job}
\address{Lockheed Martin, Sunnyvale, CA 94089, USA}
\author{Namit Anand}
\address{Quantum Artificial Intelligence Laboratory, NASA Ames Research Center, Moffett Field, CA 94035, USA}
\address{KBR, Inc., 601 Jefferson St., Houston, TX 77002, USA}
\author{Gabriel N Perdue}
\address{Superconducting Quantum Materials and Systems Center, Fermi National Accelerator Laboratory, Batavia, IL 60510, USA}
\author{Silvia Zorzetti}
\address{Superconducting Quantum Materials and Systems Center, Fermi National Accelerator Laboratory, Batavia, IL 60510, USA}
\author{M. Sohaib Alam}
\ead{malam@usra.edu}
\address{Quantum Artificial Intelligence Laboratory, NASA Ames Research Center, Moffett Field, CA 94035, USA}
\address{USRA Research Institute for Advanced Computer Science, Mountain View, CA 94043, USA}

\newpage

\begin{abstract}

High-coherence cavity resonators are excellent resources for encoding quantum information in higher-dimensional Hilbert spaces, moving beyond traditional qubit-based platforms. A natural strategy is to use the Fock basis to encode information in qudits. One can perform quantum operations on the cavity mode qudit by coupling the system to a non-linear ancillary transmon qubit. However, the performance of the cavity-transmon device is limited by the noisy transmons. It is, therefore, important to develop practical benchmarking tools for these qudit systems in an algorithm-agnostic manner. We gauge the performance of these qudit platforms using sampling tests such as the Heavy Output Generation (HOG) test as well as the linear Cross-Entropy Benchmark (XEB), by way of simulations of such a system subject to realistic dominant noise channels. We use selective number-dependent arbitrary phase and unconditional displacement gates as our universal gateset. Our results show that contemporary transmons comfortably enable controlling a few tens of Fock levels of a cavity mode. This framework allows benchmarking even higher dimensional qudits as those become accessible with improved transmons.
\end{abstract}


\section{Introduction and Motivation}
\label{secn:intro}

Quantum computers may offer significant advantages over classical machines for certain classes of problems.
Although theoretical work has shown computational complexity advantages for quantum algorithms in specific cases, significant work remains in order to realize these advantages on hardware.
Given the nascent nature of such hardware, many potential and vastly different platforms relying on a variety of physical implementations such as superconducting qubits \cite{krantz2019quantum,kjaergaard2020superconducting}, neutral atoms \cite{Wintersperger2023}, trapped ions \cite{bruzewicz2019trapped} and photonic circuits \cite{Flamini_2019,slussarenko2019photonic} are being investigated. It is not yet clear which, or even if, a particular platform will possess the correct mix of noise resilience, error correction success rate, and number of logical information carriers to efficiently execute useful quantum computing algorithms. Indeed, different hardware platforms may be matched to different problems.
Error correction and mitigation performance also varies across hardware platforms.
Just as various metrics are available to gauge classical computers' capabilities (such as CPU clock speed, I/O-bound throughput, disk storage space and monitor resolution), it is important to formulate benchmarks and metrics to gauge the performance of quantum computers so that we may better understand progress on, and differences between, various hardware implementations.

However, it is nontrivial to design a single metric that accurately captures the nebulous notion of ``performance'' across different hardware platforms.
A class of prominent tests which aims to quantify a quantum device's utility, or ``algorithmic reach'', are the volumetric benchmarks of~\cite{BlumeKohout2020volumetricframework} (see also, the associated notions of algorithmic speedup~\cite{Pokharel2023} and algorithmic error tomography~\cite{Pokharel2024}). On the one hand, a reasonable definition of algorithmic reach might be to quantify the largest instance of a specific agreed-upon algorithm that can be executed successfully on a quantum device and whose solution can be verified (likely via classical simulation). For example, what is the largest instance of a Grover search problem that one can implement on a quantum computer with a success probability greater than $2/3$? Such a definition of algorithmic reach would clearly be an \textit{algorithm dependent} metric, as opposed to other typical quantum hardware benchmarks such as the average gate fidelities that are measured via randomized benchmarking \cite{Emerson2005,Emerson2007,Knill2008,Dankert2009,Magesan2011,Helsen2022}. On the other hand, we would like to assess how well the device would perform in an \textit{algorithm agnostic} manner. This idea can be captured by benchmarking how well the device can compile and execute random unitaries, as this would be agnostic to any particular algorithm (which is typically represented, in the noiseless limit, by a single unitary).

For this purpose, we capture the notion of algorithmic reach using sampling tests such as the Heavy Output Generation (HOG) test~\cite{aaronson2016complexitytheoretic} and linear Cross-Entropy Benchmark (XEB)~\cite{QSupremacy}, applied in our case in particular to a novel quantum computing platform comprising a superconducting transmon qubit dispersively coupled to a cavity resonator mode~\cite{Reagor2016coax, Rosenblum2023Mushroom, Oriani2024Nb_cav, Yu2024stabilization}. While the inherent infinite-dimensional Hilbert space of the cavity mode allows different bases and schemes in which to encode quantum information, a simple choice is to utilize its individual energy eigenstates, such that it naturally forms a higher-dimensional qudit. High energy physics theorists have explored algorithms for such devices \cite{alam2022quantumcomputinghardwarehep, kurkcuoglu2022quantumsimulationphi4theories, gustafson2022primitive, gustafson2021large, gustafson2021prospects, illa2024qu8its, gonzalez2022hardware}. On its own, the cavity mode would function as a harmonic oscillator with equally spaced energy levels. This makes it impossible to address its individual energy levels, rendering the cavity incapable of universal quantum computation. However, by coupling it to a non-linear anharmonic oscillator in the form of a transmon, one can overcome this limitation. Indeed, a similar architecture based on a hybrid oscillator-qubit system was recently studied in Ref.~\cite{liu2024hybridoscillatorqubit}.

We constrain our study to a full realistic simulation of a single cavity mode \cite{Roy2024Qudit}. While the development of a multi-mode system is an active field of research in the community~\cite{Chakram2021seamless, Reineri2023multimode, you2024crosstalk, Roy2023multimode}, the optimal configuration for, and operation of, a multi-qudit computer is not yet fully clear, in contrast to qubit based platforms which have been more widely studied. 

However, much can still be gleaned from such a single-mode system and applied to future versions. In the current NISQ-era of quantum computing, real hardware is subject to uncontrollable noise sources of various physical origins, which have deleterious effects. Understanding how such a device performs under benchmarking tests for both various noise strengths as well as different theoretically-desired parameters, such as the qudit's dimension (which encapsulates its quantum information storage and processing potential), allows for the identification of bottlenecks and serves to guide future research and engineering efforts. In particular, such tests are easily repeatable as the dimension of the qudit is scaled, and could be seen as an indication of how easy it is to control such a device. Furthermore, powerful quantum computing devices will inexorably depend on highly performant classical computational efforts. This is so, given the necessity of classical optimizers and compilers to efficiently translate higher-level algorithms down to hardware-level instructions, as well as the need for fast classical decoding algorithms for quantum error correction.

Our goal in this work is to study algorithmic-complexity-inspired benchmarks with which to evaluate the progress in both hardware and software development for single-qudit devices at present, in order to lay the foundation for benchmarking multi-qudit devices in the future. The two aforementioned metrics, namely the HOG test (developed initially for studying the performance of qubit-based quantum computers~\cite{cross2019validating}) and XEB (which was initially developed as a potential demonstrator calculation for quantum advantage and is now understood to have certain drawbacks \cite{gao2024limitations}, but which still serves as a general metric for quantum computational performance) will be applied to such a system.

The rest of the paper is organized as follows. In Sec. \ref{secn:physical} we describe, in detail, the physical cavity-qubit system, its universal control down to the pulse level for our chosen gateset, as well as its dynamics under a realistic noise model. Sec. \ref{sec:metrics} discusses the two useful metrics we use to benchmark this system. The numerical results of this full-stack simulation are then presented and discussed in Sec. \ref{secn:results}, and we finally conclude with the future outlook in Sec. \ref{secn:conclusion}.

\section{Physical System}
\label{secn:physical}

\subsection{Theory}
\label{subsecn:theory}

The areas of cavity and circuit quantum electrodynamics are extremely rich fields in their own rights, and we direct the interested reader to \cite{blais2021circuit} for a comprehensive introduction to the topics. The model under consideration consists of a transmon qubit dispersively coupled to a single resonator mode, each with independent drives, in the laboratory frame
\begin{align}
\hat{H} = \hat{H}_{\text{disp}} + \hat{H}_d,
\end{align}

\noindent
where

\begin{align}
\hat{H}_{\text{disp}} & = \omega_{a}a^{\dagger}a + \omega_{q}\ketbra{e} + \chi a^{\dagger}a \ketbra{e}, \nonumber \\
\hat{H}_d & = \epsilon_1(t) \left[ e^{-i(\omega_1t + \phi_1)}a^{\dagger} + e^{i(\omega_1t + \phi_1)}a \right] \nonumber \\
& + \epsilon_2(t) \left[ e^{-i(\omega_2t + \phi_2)}\sigma^{+} + e^{i(\omega_2t + \phi_2)}\sigma^{-} \right].
\end{align}

\begin{figure}[ht]
\centering
\includegraphics[width=0.7\linewidth]{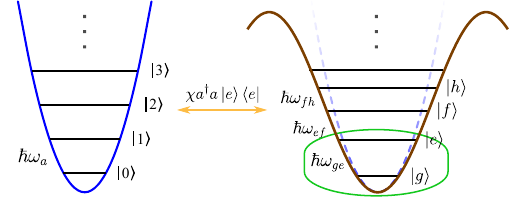}
\caption{A single linear oscillator cavity mode coupled to a non-linear transmon qubit (with $\omega_{ge} = \omega_{q}$) in the dispersive regime.}
\label{fig:system}
\end{figure}

Here, $\omega_{a}$ and $\omega_{q}$ are the \textit{dressed} frequencies of the resonator mode with annihilation operator $a$ and of the qubit with excited state $\ket{e}$. The dispersive shift is denoted by $\chi$. Furthermore, $\epsilon_{1(2)}$ is the real time-dependent amplitude, and $\phi_{1(2)}$ a time-dependent phase, of the cavity(qubit) drive with frequency $\omega_{1(2)}$\footnote{\label{ftn:drive}A general drive, in terms of its in-phase $I$ and out-of-phase $Q$ quadratures, can be expressed as $I(t)\cos(\omega_dt) + Q(t)\sin(\omega_dt)$. This is equivalent to the expression $\epsilon_d(t) \cos(\omega_dt + \phi_d)$, with $\epsilon_d(t)$ the drive amplitude, $\phi_d$ the drive phase and $\omega_d$ the carrier frequency.}.
This Hamiltonian is only valid in the dispersive regime whereby the qubit-cavity detuning, $\Delta = \omega_q - \omega_{a}$, is large with respect to the resonator-qubit coupling $g$, i.e. $g \ll |\Delta|$.
See Fig.~\ref{fig:system} for a schematic cartoon of the setup.

Moving into a frame rotating at both the cavity and qubit frequencies (using the time-dependent unitary transformation $\hat{U}=e^{i\omega_aa^{\dagger}at}e^{i \omega_q \ketbra{e}t}$), the Hamiltonian becomes

\begin{align}
\hat{H}^{'} & = \hat{U} \hat{H} \hat{U}^{\dagger} + i \dot{\hat{U}} \hat{U}^{\dagger} = \hat{H}_{int} + \hat{H}_{1}(t) + \hat{H}_{2}(t) \nonumber \\
& = \chi a^{\dagger}a \ketbra{e} + \epsilon_1(t) \left[ e^{-i(\delta_{1a}t+\phi_1)}a^{\dagger} + \text{h.c.} \right] \nonumber \\
& + \epsilon_2(t) \left[ e^{-i(\delta_{2q}t+\phi_2)}\sigma^{+} + \text{h.c.} \right].
\label{eqn:hamiltonian}
\end{align}

Here $\delta_{1a} = \omega_1 - \omega_a$ and $\delta_{2q} = \omega_2 - \omega_q$ are the detunings of the drive tones from the resonator and qubit frequencies, respectively. Next, notice that, for the usual displacement operator $\hat{D}(\alpha)$ given by

\begin{equation}
\hat{D}(\alpha) = e^{\alpha a^{\dagger} - \alpha^{*}a},
\label{eqn:displacement}
\end{equation}

\noindent
the cavity drive Hamiltonian $\hat{H}_1(t)$ is $\hat{D}$'s Lie algebra generator. Furthermore, the qubit drive $\hat{H}_2(t)$ effects a transition between its ground and excited states, with the particular dynamics controlled by tuning both $\phi_2$ and $\epsilon_2(t)$.

The computational subspace of the qudit is hosted in the lowest $d$ energy eigenstates, known as the Fock states, of the resonator. Even though $\hat{D}(\alpha)$ acts on an infinite dimensional Hilbert space, it is possible to capture its effect using a carefully chosen resonator dimension with minimal error as discussed later. 
As we shall see, the qubit intermediary is important for creating Selective Number-dependent Arbitrary Phase (SNAP) pulses \cite{SNAP2015PRA, SNAP2015PRL}, which allow one to selectively add arbitrary phases to specific Fock states of the cavity. A successful SNAP pulse is parameterised by the angles $\vec{\theta} = (\theta_0, \theta_1, \cdots, \theta_{d-1})$ and operates, with a high fidelity, on the first $d$ Fock states of the cavity via $\hat{S}(\vec{\theta})$, where

\begin{equation}
\hat{S}(\vec{\theta}) = 
\begin{pmatrix}
e^{i\theta_{0}} & 0 & \cdots & 0 \\
0 & e^{i\theta_{1}} & \dots & 0 \\
\vdots & \vdots & \ddots & \vdots \\
0 & 0 & \dots & e^{i\theta_{d-1}}
\end{pmatrix}.
\label{eqn:snapmatrix}
\end{equation}

A general closed quantum system's evolution $\hat{U}(t, t_0)$, from an initial time $t_0$ to some later time $t$, is given by

\begin{align}
\hat{U}(t &, t_0) = \mathcal{T} e^{-i \int_{t_0}^t \hat{H}^{'}(\tau)d\tau} = \sum_{n=0}^\infty \hat{U}_n(t, t_0) \nonumber \\
& = \sum_{n=0}^\infty \frac{(-i)^n}{n!} \int_{t_0}^t dt_1 \cdots \int_{t_0}^t dt_n \mathcal{T} \left[ \hat{H}^{'}(t_1) \cdots \hat{H}^{'}(t_n) \right],
\label{eqn:dyson}
\end{align}

\noindent
with $\mathcal{T}$ the time-ordering operator. The time dependence of the drives complicates finding a general closed-form expression for $\hat{U}$ explicitly. However, it has been shown that a combination of displacement and SNAP pulses provides universal control over the resonator cavity~\cite{SNAP2015PRA,SNAP2015PRL}. As such, we use the following ansatz for the unitary evolution of our system

\begin{equation}
\hat{U} \approx \hat{D}(\alpha_{k+1}) \hat{S}(\overline{\theta}_k) \hat{D}(\alpha_{k}) \cdots \hat{S}(\overline{\theta}_{1}) \hat{D}(\alpha_{1}),
\label{eqn:snapdisplacementansatz}
\end{equation}

\noindent
where $k$ is the number of ``SNAP-displacement'' layers in the ansatz (note that there are $k+1$ displacements). As a rough guess of the requisite number of layers, first note that for a qudit computational subspace of dimension $d$, each SNAP+displacement layer in Eq.~\eqref{eqn:snapdisplacementansatz} has $d+1$ real parameters. Next, although any arbitrary $SU(d)$ unitary can be expressed exactly with an ansatz comprised of $O(d^2)$ many SNAP+displacement layers, a dimensional counting argument~\cite{fösel2020efficient} implies that it may also be possible to make use of an $O(d)$ length ansatz. In particular, the argument is that since the $SU(d)$ real Lie group has dimension $d^2-1$ (i.e. any $SU(d)$ matrix has $d^2-1$ real parameters), and each SNAP+displacement layer possesses $d+1=O(d)$ independent real parameters, an ansatz comprised of an $O(d)$ number of layers possesses sufficient real parameters to fit those of the $SU(d)$ matrix being compiled. Such an ansatz is therefore expressive enough for one to be able to compile arbitrary unitaries, given a sufficiently large depth or number of layers \cite{SNAP2015PRA}. However, it suffers from a barren plateau problem in terms of its trainability~\cite{ogunkoya2023investigating}. Note that compiling an arbitrary $SU(d)$ unitary onto qubit based platforms also generically requires $O(d^2)$ 2-qubit gates \cite{barenco1995elementary, knill1995approximation}, and that variational ansatzes designed to compile Haar random unitaries also suffer from barren plateau problems~\cite{mcclean2018barren}.

In practice, we limit the number of layers in the ansatz to as small a number as possible, in order to avoid barren plateaus and maintain trainability.
A similar counting argument suggests that since an arbitrary $d$-dimensional pure state possesses $2d - 1 = O(d)$ real parameters, it may be possible to prepare such a state (i.e. given some $d$-dimensional target state $\ket{\psi}$, find $U$ such that $\ket{\psi} = U \ket{0}$), using a constant, i.e. $O(1)$, number of SNAP+displacement layers. As we'll see below, our metrics rely on sampling from certain probability distributions, which in turn reduce to the problem of state preparation, which is simpler than unitary compilation.

We next describe the displacement and SNAP pulses in detail to make the mapping from the gate-level parameters to physical pulse profiles clearer.

\subsection{Displacement pulse}
\label{subsec:displacement}

The displacement operator $\hat{D}(\alpha)$ in Eq.~\ref{eqn:displacement} describes the displacement of an arbitrary state in phase space by an amount $\alpha$, and creates coherent states $\ket{\alpha}$ by displacing the vacuum state from the origin, i.e. $\ket{\alpha} = \hat{D}(\alpha) \ket{0}$. To map this operator to the pulse level, first assume that the resonator and qubit drive pulses are never applied simultaneously ($\epsilon_1(t) \neq 0 \implies \epsilon_2(t) \equiv 0$, and vice-versa). To create a cavity displacement pulse during $t_1$ to $t_2$, we need to ensure that whenever the cavity is driven, the qubit is in its ground state. In this case, only the $\hat{H}_1(t)$ term in Eq.~\ref{eqn:hamiltonian} survives. Driving the cavity on resonance ($\delta_{1a} = 0$) and setting $\phi_1$ to be constant throughout the duration of the displacement pulse, $\hat{H}_1(t)$ commutes with itself for all times. In this case, we can drop $\mathcal{T}$ in Eq.~\ref{eqn:dyson} so that

\begin{equation}
\hat{U}(t_2, t_1) = e^{-i\int_{t_1}^{t_2} \epsilon_1(\tau) \left( e^{-i\phi_1}a^{\dagger} + e^{i\phi_1}a \right) d\tau} \equiv \hat{D}(\alpha),
\end{equation}

\noindent
where $\alpha = -i\int_{t_1}^{t_2} \epsilon_1(\tau)e^{-i\phi_1}d\tau$. Choosing the pulse envelope $\epsilon_1(t)$ with compact support on the domain $\left[ t_1, t_2 \right]$ and constant phase $\phi_1$ allows us to tailor an arbitrary phase space cavity displacement pulse.
Note that although the number operator $a^{\dagger}a$ has countably infinite eigenvalues in the cavity Fock basis, in numerical simulations we truncate the cavity Hilbert space at some finite cutoff $N_{\rm cavity}$ that is chosen to be much larger than the qudit dimension $d$, such that it is near, but conservatively above, the minimum value at which the computation is sensitive to the cutoff level.

\subsection{SNAP pulse}
\label{subsec:SNAP}

The SNAP \cite{SNAP2015PRL} unitary in Eq.~\ref{eqn:snapmatrix} is realized through a phase kick-back on the cavity mode coupled to the qubit. First, at the beginning of a SNAP gate, we ensure that the qubit is in its ground state. At this stage, we assume a generic superposition state $\ket{\psi}_c = \sum_{n=0}^{\infty}c_n \ket{n}$ for the cavity, and an initial separable composite state of the system, i.e. $\ket{\psi}_c\ket{g}$. Next, recall that a rotation of the qubit state about an arbitrary axis $\hat{m} = (m_x, m_y, m_z)$ by angle $\theta$ on the Bloch sphere is described by

\begin{align}
R&_{\hat{m}}(\theta) = e^{-i\theta \hat{m} \cdot \vec{\sigma}/2} \nonumber \\
& = \cos{\left( \frac{\theta}{2} \right)}I - i\sin{\left( \frac{\theta}{2} \right)}\left( m_x \sigma_x + m_y \sigma_y + m_z \sigma_z \right).
\end{align}

\begin{figure}
    \centering
    \begin{quantikz}[row sep={0.8cm,between origins}]
        \lstick{$\ket{\psi}_c$} & \ctrl{1} & \ctrl{1} & \qw \\
        \lstick{$\ket{0}$} & \gate[]{R_{\hat{y} | n}(\pi)} & \gate[]{R_{ \hat{\theta}'_n | n}(\pi)} & \qw \\
    \end{quantikz}
    \caption{SNAP gate realization through the application of two $\pi$ pulses on a qubit, conditioned on the cavity $\ket{n}$ state, about axes $\hat{y} = (0,1,0)$ and $\hat{\theta}'_n = (-\sin(\theta_n'), \cos(\theta_n'), 0)$ with $\theta_n'$ the anti-clockwise angle between the Bloch sphere $\hat{y}$ and $\hat{\theta}_n$ axes.}
    \label{fig:snap_pi_pulses}
\end{figure}
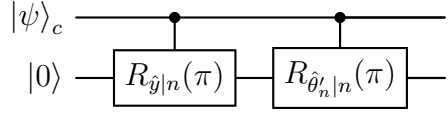

To add a geometric phase $e^{i\theta_n}$ to Fock state $\ket{n}$ using an ideal SNAP sequence, we apply two $\theta=\pi$ pulses to the qubit (setting $m_z = 0$ from here onward), conditioned on the cavity state $\ket{n}$, with the horizontal rotation axis of the second $\pi$ pulse advanced by an angle $\theta_n' = \pi-\theta_n$. This is depicted in Fig. \ref{fig:snap_pi_pulses}, with the two operators given by

\begin{align}
R_{\hat{y} | n}(\pi) & = \ketbra{n} \otimes R_{\hat{y}}(\pi) + \sum_{n' \neq n} \ketbra{n'} \otimes I, \\
R_{\hat{\theta}'_n | n}(\pi) & = \ketbra{n} \otimes R_{\hat{\theta}'_n}(\pi) + \sum_{n' \neq n} \ketbra{n'} \otimes I.
\end{align}

It can be shown that these two operators create the desired cavity-qubit state

\begin{equation}
\sum_{n=0}^{\infty}c_n \ket{n}\ket{g} \to (e^{i\theta_n}c_n \ket{n} + \sum_{n' \neq n}c_{n'} \ket{n'})\ket{g},
\end{equation}

\noindent
thereby realising a single SNAP angle $S_n(\theta_n) = \exp(i\theta_n \ketbra{n})$. 

To compute the pulse level representation of a SNAP gate, we first assume that the cavity drive is turned off. In this case the system Hamiltonian becomes $\hat{H}^{'} = \hat{H}_{\rm int} + \hat{H}_{2}(t)$. In the current cavity-qubit co-rotating frame, all basis states with the qubit in the ground state, $\{ \ket{n, g} \}_{n=0}^{\infty}$, have resonance frequencies of $0$. The frequencies of the $\{ \ket{n, e} \}_{n=0}^{\infty}$ states, however, are all respectively $n\chi$. To add an arbitrary relative phase $\theta_n$ to a \textit{single} chosen Fock state $\ket{n}$ of the resonator, we tailor the first $\pi$ pulse to have a carrier frequency of $\delta_{2q} = n\chi$ and hence drive the transition $\ket{n, g} \leftrightarrow \ket{n, e}$.
Calibrating the pulse entails choosing $\epsilon_{2}(t)$ between times $t_1$ and $t_2$ (usually a flat-top or truncated Gaussian envelope) such that $\int_{t_1}^{t_2}\epsilon_{2}(t)dt = \pi$. For this first pulse we set $\phi_{2} = 0$ without loss of generality, which amounts to fixing the horizontal Bloch sphere axis about which we rotate the qubit (arbitrarily chosen to be the $\hat{y}$ axis in Fig.~\ref{fig:snap_pi_pulses}).
We then apply a second $\pi$ pulse with identical envelope and drive frequency, but with a drive phase of $\phi_{2} = \pi -\theta_n$. This rotates the state $\ket{n, e}$ back to the qubit ground state $\ket{n, g}$, but now about an axis that is offset by $\pi - \theta_n$ with respect to the first (see Figs.~1 in both Refs.~\cite{SNAP2015PRA} and \cite{SNAP2015PRL} for intuitive illustrations of this). The net effect is that the qubit trajectory causes the state $\ket{n, g}$ to acquire a geometric phase $e^{i\theta_n}$, as desired, while leaving the qubit and cavity disentangled.

We have yet to consider the effect of this single SNAP operation on the other basis states in $\ket{\psi}_c\ket{g}$. The energy of $\ket{n, e}$, namely $n\chi$, is separated from that of its neighboring states in the co-rotating frame by $\chi$. While driving $\ket{n, g} \leftrightarrow \ket{n, e}$, we wish to minimize other transitions $\ket{m, g} \leftrightarrow \ket{m, e}$ for $m \neq n$. This is done by ensuring that the spectral linewidth of $\epsilon_{2}$, which is roughly proportional to its peak height $|\epsilon_{2}|$ in the time domain profile, is much smaller than the spacing between adjacent transitions, i.e. $|\epsilon_{2}| \ll |\chi|$. This condition imposes a fundamental trade-off in SNAP pulses: to ensure this weak drive condition while maintaining a pulse area of $\pi$, drive $\epsilon_{2}$ must be longer in time. This is the reason why SNAP pulses are relatively long in practice. For a single SNAP pulse conditioned on a single $n$, we choose $|\epsilon_{2}| = \chi/10$. Consequently, a longer $\pi$ pulse causes the qubit to spend more time in the excited state, making the SNAP gate more susceptible to qubit decay and dephasing noise.

A generic multi-component SNAP gate $\hat{S}(\vec{\theta}) \equiv \Pi_{j=0}^{d-1} S_j(\theta_j)$ in Eq.~\eqref{eqn:snapmatrix} can be realized by applying multiple such conditional qubit rotations either sequentially or simultaneously, where the latter is preferred to reduce gate time. One can add arbitrary phases to the first $d$ Fock levels of the cavity by multiplexing the qubit drive signal such that the first pulse of the SNAP sequence is the sum of $d$ waveforms each with the same amplitude $\epsilon_2(t)$ and drive phases of $\phi_2 = 0$, but each with a respective carrier frequency of $j \chi$ for all $j = 0, \cdots, d-1$. As such, each component drives a transition $\ket{j, g} \to \ket{j, e}$. The second pulse is identical except with the phases set to the respective desired values $\theta'_j$. Ideally, this would allow us to create the unitary in Eq.~\eqref{eqn:snapmatrix} with high fidelity. However, there may be practical limitations to this scheme as discussed below. 

In the best case scenario, as $d$ increases, the SNAP gate length (which is twice the length of each $\epsilon_2(t)$) can be kept independent of $d$. Let $P_{\text{s}}$ be the microwave power delivered to the system while implementing a constituent $S_j(\theta_j)$, taking time $t_s=2(t_2-t_1)$ (with $P_{\text{s}} \propto p = \int_{t_1}^{t_2} |\epsilon_2(t)|^2dt$).
The total power delivered for a SNAP gate would hence be $P_{\text{s}}d$, which for large $d$ would heat the system and cause noise and other unwanted effects, e.g. AC-Stark shifts~\cite{carroll2022dynamics}. One could apply the $S_j(\theta_j)$s sequentially to avoid any heating, but at the expense of increasing the total gate time to $t_sd$. This in turn would result in reduced system performance due to decoherence. We employ a trade-off via the simultaneous application of $S_j(\theta_j)$s with longer gate times, while limiting the total delivered power. 

To understand the scaling with $d$, let $P_{\text{max}} = P_{\text{s}}$ be the maximum power we allow to be delivered to the system, and let $P'_{\text{s}}$ be the new modified power for each constituent $S_j(\theta_j)$ with $P'_{\text{s}} = P_{\text{s}}/d$. Since the Rabi rate is proportional to $\sqrt{P'_{\text{s}}}$, in order to compensate for the lower rate, the gate time of the $d$-dimensional SNAP gate needs to be $t_s\sqrt{d}$. Therefore, we scale the length of a $d$-dimensional SNAP pulse with the square root of $d$ in our simulations.

\subsection{Noise model}
\label{subsec:noisemodel}

We consider various benchmarking metrics for our resonator-transmon system in the presence of noise~\cite{Roy2024Qudit}. While $\hat{H}^{'}$ generates the unitary dynamics, common incoherent sources of noise are generally described by non-unitary operators and the dynamics consequently described by a more general master equation. For resonator cavities, the dominant noise channel is generally photon loss. Contemporary 3D cavities typically have very high $Q$ values (a measure of energy loss) exceeding $10^9$ \cite{Romanenko2020Tesla, Oriani2024Nb_cav}. On the other hand, superconducting transmon qubits typically show  $Q$ values around $10^6$~\cite{Burnett2019, carroll2022quantum, kurter2022quasiparticle}, but are nevertheless necessary to couple to the cavity in order to drive certain gates such as the SNAP gate. Hence we can safely ignore losses intrinsic to the cavity and assume that errors on the gates are primarily caused by the losses in the transmon. The coupling of transmons to the noisy environment, either represented as a bath and/or two-level systems \cite{Mueller_2019, Cho_2023}, once the environment is traced out, is best modeled as non-unitary qubit decay and dephasing channels.

Under the Born-Markov approximation \cite{breuerTheory2007}, assuming that our system and environment are initially uncorrelated, the master equation describing the cavity-qubit dynamics takes the Lindblad form

\begin{equation}
\dot{\rho} = -i\left[ \hat{H}^{'}, \rho \right] + \gamma_1 \mathcal{D}[\sigma^{-}]\rho + 2\gamma_{\phi} \mathcal{D}[\sigma_z]\rho,
\label{eqn:lindblad}
\end{equation}

\noindent
where the collapse operators are $\sqrt{\gamma_1}\sigma^{-}$ and $\sqrt{\gamma_{\phi}}\sigma_z$.
In this expression, $\gamma_1 = 1/\mathrm{T}_1$ is the qubit decay rate, $\gamma_{\phi} = 1/\mathrm{T}_2 - 1/\left(2\mathrm{T}_1\right)$ is the pure dephasing rate, and $\sigma^{-}$ and $\sigma_z$ are the system-environment coupling operators. The superoperator $\mathcal{D}$ is defined as $\mathcal{D}[b]\boldsymbol{\cdot} = b \boldsymbol{\cdot} b^{\dagger} - \frac{1}{2} \{ b^{\dagger}b, \boldsymbol{\cdot} \}$. We used \texttt{QuTiP} \cite{johansson20121760, johansson20131234} to simulate the system dynamics.

\section{Metrics}
\label{sec:metrics}

Currently, there is no universally agreed upon standard test suite to benchmark the algorithmic performance of quantum computers using different hardware paradigms.
Moreover, existing metrics are often difficult to compute in practice, due to the exponential scaling of the Hilbert space dimension with the number of qubit/qudit information carriers.

An example of an ideal metric would be to measure a device's fidelity, i.e. $\mathcal{F} = \bra{\psi} \rho \ket{\psi}$, with $\rho$ being the noisy implementation of an ideal error-free state $\ket{\psi}$ that one may be interested in preparing. However, this technique generally requires some form of quantum tomography, which is prohibitively expensive for large Hilbert space sizes. Although recently developed shadow tomography techniques \cite{aaronson2018shadowtomographyquantumstates, Huang2020} can bring this cost lower, it is not clear how to adapt these techniques to qudit based platforms such as the one we consider here. Part of this difficulty lies in the nonexistence of unitary \(2\)-groups (unitary \(2\)-designs with an additional group structure) in arbitrary dimensions \cite{bannai_unitary_2018}, as well as in generalizing Gottesman-Knill type results to arbitrary \(d\) that is not prime or power of a prime~\cite{gottesman1998fault}.

Moreover, we are primarily interested in the performance of the device as we use it for increasingly complex unitary operations. Most benchmarking efforts in qubit-based platforms have focused on sampling problems. The chosen metrics allow a relatively small number of measurements on a quantum device prepared in a state representing a specific probability distribution to be taken while still providing some approximation of a robust underlying ``quantum'' metric. Two such empirical metrics are \textit{quantum volume} \cite{Moll_2018,cross2019validating} and the \textit{cross-entropy} benchmark \cite{BoixoXEB,QSupremacy}.

The quantum volume of a quantum processor unit (QPU) is a performance metric that quantifies the maximum size of a ``square'' quantum circuit that can be successfully run on a QPU. It is a ``full stack'' metric, in that it is a single number that captures many real-world features and imperfections of a QPU. It takes into account the qubit/qudit and gate quality, qubit/qudit connectivity, classical compiler performance, and coherent and incoherent noise in the electronics and devices themselves, among other potential sources of imperfections.

When benchmarking a single qudit, the quantum volume ansatz circuit reduces to applying a Haar random unitary on that qudit; there is no need for the SWAP operations since there is only one information carrier. This is analogous to choosing random $SU(2)$ unitaries on individual qubits in the qubit-based quantum volume setting. In part due to this reason, it is difficult to define the notion of the quantum volume of a single qudit. However, in this case employing sampling tests such as the heavy output generation (HOG) test \cite{aaronson2016complexitytheoretic} or the linear cross-entropy benchmark (XEB) \cite{QSupremacy,ware2023sharp}, described in greater detail in the following sub-sections, still makes sense.
Note that in the case of multiple qudits, with the addition of multi-qudit gates including possibly SWAP operations, a straightforward modification of quantum volume may again become a good benchmark \cite{BlumeKohout2020volumetricframework}.

In the quantum setting, given that we typically perform repeated single-shot measurements in a single basis (usually the $\sigma_z$ basis for qubits and the Fock basis in the current cavity QED context), we are in fact measuring a classical distribution derived from a quantum state. As such, focusing on the cross-entropy between $Q$ (derived from the noiseless state $\ket{\psi}$) and $P$ (derived from the noisy state $\rho$) makes intuitive sense. However, two points are worth mentioning. First, studies typically consider a linear version of $H$, namely XEB, to better minimize fluctuations when empirically measuring entropy \cite{QSupremacy, gao2024limitations}. The XEB of a device is typically taken over an ensemble of random circuit implementations, in order to gauge the performance of a QPU itself rather than for a single unitary. And second, the XEB is meant to serve as a stand-in for the circuit fidelity of the QPU \cite{QSupremacy, BoixoXEB, choi2023preparing}. Below, we outline the details of each metric.

\subsection{Heavy output generation test}
\label{subsecn:hog}

Within the context of qubit-based systems, the most common quantum volume scheme \cite{cross2019validating} entails first choosing a small subset of $m$ of the available qubits. The action of a large $2^m \times 2^m$ Haar random unitary $U$ acting on the subset is to be approximated by performing Haar random two-qubit unitary operations between randomly selected pairs of qubits from the subset.
This creates a single ``layer''; such layers are concatenated together until the number of layers, $d_{\ell}$ equals the number of qubits ($m$).
This defines a ``square'' ansatz circuit.
The choice of a square circuit --- namely settings $d_{\ell} = m$ --- was motivated by the difficulty of simulating random circuits of equal width and depth using tensor network arguments \cite{cross2019validating}, although relaxing this restriction gives rise to other volumetric benchmarks \cite{BlumeKohout2020volumetricframework}.

Measurements of the state of each qubit are then performed, with each measurement producing a bitstring $x \in \{ 0,1 \}^m$. For an \textit{ideal} QPU and a single unitary $U$, the ideal output bitstring distribution is defined by

\begin{equation}
q_U\left(x\right) = \vert \langle x \vert U \vert 0 \rangle \vert^2.
\label{eqn:idealbitstringprob}
\end{equation}

From this ideal distribution, the so-called heavy output set corresponding to $U$ is computed, as
\begin{equation}
H_U = \{ x \in \{0,1\}^m \text{~such~that~} q_U\left(x\right) > q_{\text{med}}\},
\label{eqn:hogheavyset}
\end{equation}
\noindent
where $q_{\text{med}}$ is the median bitstring probability from the set $\{ q_U\left(x\right) \}$.

In reality, $q_U$ is not the distribution that is sampled from. The unitary $U$ is classically compiled into the square ansatz circuit discussed above, implemented on real hardware, and the state of each qubit read out. Various sources of noise and imperfections compound in such a chain such that the actual distribution that is sampled from is not that implied by $U$, but rather a noisy $U'$ ($U'$ is almost always in fact a non-unitary noisy quantum channel).

This leads us to the heavy output generation (HOG) test \cite{aaronson2016complexitytheoretic}, an important subroutine in quantum volume benchmarking. 
For the single $U$, the multiple noisy quantum bitstring measurements are classified as lying in their corresponding classically-computed heavy set $H_U$ or not.
This entire chain - from generating a random $U$, to computing the heavy set, compiling the physical implementation of $U$ according to the square ansatz and collecting bitstring statistics from multiple physical measurements - is repeated for a large ensemble of Haar random unitaries.
The HOG test finally entails examining all of the output bitstring measurements from the ensemble of unitaries and, for the circuit size $m$, defining ``success'' as the case where more than two-thirds of the bitstring measurements lie in their respective heavy set\footnote{\label{fnt:hogfractions}It is known that this HOG fraction metric falls to $0.5$ for a completely depolarized QPU (which would be akin to sampling from uniformly random noise) and asymptotically approaches $\left(1 + \ln 2\right)/2 \simeq 0.85$ for a perfect QPU \cite{cross2019validating}.}.

The HOG subroutine test is mapped to a single numeric metric for the system: if we relax the condition that the circuit ansatz be square, i.e. not require that the depth $d_{\ell}(m)$ equal the width $m$, the quantum volume of a qubit-based system, $V_Q$ is

\begin{equation}
\log_2 V_Q = \text{argmax}_m \text{min} \left(m, d_{\ell}(m)\right),
\end{equation}

\noindent
where $\text{argmax}_m$ here denotes the largest value of $m$ for which the system passed the HOG test. In practice, one chooses a small initial value for $m$ and increments it (by adding both a single qubit and a single layer to the circuit ansatz in the case of $d_{\ell} \equiv m$) until the test fails.

It is unclear how to map the HOG test to a single number representative of the qudit device under consideration in direct analogy to quantum volume because we are not entangling multiple qudits.
As such, for our single qudit of dimension $d$, the bitstrings $x$ will now lie in the set $\{ 0, 1, \cdots, d-1 \}$, $U$ denotes a set of $d \times d$ Haar random unitaries acting on said qudit, and the HOG test itself still requires a two-third threshold.

\subsection{Linear cross-entropy benchmarking}
\label{subsecn:xeb}

Cross-entropy benchmarking is a protocol that relies on running random quantum circuits on a quantum device, which is assumed to be noisy, obtaining samples, and classically computing the ideal sampling probabilities of the outcomes. Ordinary cross-entropy, a concept fundamental to information theory \cite{cover2006element}, is a way of quantifying the similarity or difference between two classical distributions. It is defined as $H(P, Q) = -\sum_{x} P(x) \log{Q(x)}$ for distributions $P$ and $Q$. One way of intuitively understanding cross-entropy is, if one optimizes a code scheme for a model distribution $Q$, but the true underlying distribution is in fact $P$, the cross-entropy quantifies the average amount of information needed to describe an element $x$, sampled from $P$, using $Q$'s encoding scheme. While XEB has been used in the past as a protocol to demonstrate quantum supremacy \cite{QSupremacy,liu2021redefining}, here we employ it to assess how well we are able to control and manipulate the Hilbert space of a single qudit of some given dimension.

The linear cross-entropy benchmark (XEB) involves computing the linear version of $H$ from Sec. \ref{sec:metrics}, which, for the case of a qudit of dimension $d$, is

\begin{equation}
\text{XEB} = d \, \mathbb{E}_{U} \left[ p_{U}(x) \cdot q_{U}(x)\right] - 1,
\label{eqn:xeb-linear}
\end{equation}

\noindent
where $q_{U}(x)$ denotes the ideal classical bitstring distribution (derived from the pure target state $\ket{\psi}$), $p_{U}(x)$ the noisy classical distribution (derived from $\rho$ with $x \in \{ 0, 1, \cdots, d-1 \}$), and the center dot denotes the inner product. In a typical experiment, the expected value $\mathbb{E}_{U} \left[ p_{U}(x) \cdot q_{U}(x)\right]$ is estimated by sampling finitely many bitstrings for each random circuit $U$ to construct an estimator of $p_U(x)$ (using, for example, Bayesian inference, since we do not have full knowledge of $p_U$ in a lab - see Appendix Sec. \ref{secn:stats}) and computing the sample mean over many $U$'s.

In our numerical simulation, where we are interested in benchmarking the ``controllability'' of a single qudit of increasing dimension using the SNAP+displacement gateset in the presence of noise, we can easily compute the exact dot product of the probability vectors.
Moreover, whereas the ensemble of unitaries is typically taken to be random quantum circuits built out of primitive one- and two-qubit gates on qubits, here we explicitly sample $d \times d$ Haar random unitaries and compile them into the native gateset. Since the XEB metric relies on the ideal classical probability vector $q_U$ and since we assume that our qudit is always initialized in the ground state $\ket{0}$, we compile instead the shortest SNAP+displacement sequence that could create the state $U \ket{0}$ with high fidelity. As discussed previously in Sec. \ref{subsecn:theory}, Since we are concerned with state preparation here, we can limit ourselves to an $O(1)$ ansatz length. Concretely, we utilize two SNAP gates interspersed with three displacements as the ansatz, so $k=2$.

For an ideal device where $p_U = q_U$, $\mathbb{E}_{U} \left[ p_{U}(x) \cdot q_{U}(x)\right] = \frac{2}{d+1} \approx \frac{2}{d}$ for a Haar random ensemble~\cite{QSupremacy}, with $p_U$ and $q_U$ functions of the sampled $U$. If $p$ and $q$ are instead uniformly distributed, whereby each outcome $x$ is equally likely, the average $\mathbb{E}_{U} \left[ p_{U}(x) \cdot q_{U}(x)\right] \approx \frac{1}{d}$ instead~\cite{QSupremacy}.
This produces the two limits of $\text{XEB} \approx 1$ for an ideal quantum device with very large $d$, and $\text{XEB} \approx 0$ for a uniform sampler. This latter case is easy to simulate classically by uniformly sampling random bitstrings.

There is complexity-theoretic evidence that exactly sampling from the output of random quantum circuits is classically difficult and remains so even in the presence of a small amount of noise \cite{Bouland2019}. This implies that achieving $\text{XEB} \approx 1$ is likely infeasible for any classical algorithm.
For our purposes however, we employ the XEB score to benchmark how reliably we are able to compile and affect Haar random unitaries on our cavity mode qudit, without the aspiration of demonstrating any form of supremacy.
This is because, at least in the single qudit case, the entire protocol is easily classically simulable with a runtime that scales polynomially in the dimension $d$ of the qudit.

Finally, for relatively small values of $d$ such as the ones we explore here, the value for $\text{XEB}$ can be substantially smaller than 1. As such, we instead consider a normalized version of $\text{XEB}$, denoted $\text{XEB}_n$, as proposed in \cite{dalzell2021random} and employed in \cite{ware2023sharp}. It is defined as
\begin{equation}
\text{XEB}_n = \frac{ \text{XEB}}{ d \, \mathbb{E}_{U} \left[ q_{U}(x) \cdot q_{U}(x)\right] - 1}.
\label{eqn:xeb-normalized}
\end{equation}
This quantity lies between $0$ and $1$, assuming the former value in the case of a depolarized QPU and the latter when sampling from a perfect one.

\section{Results}
\label{secn:results}

\begin{figure}[t]
\centering
\includegraphics[width=0.7\linewidth]{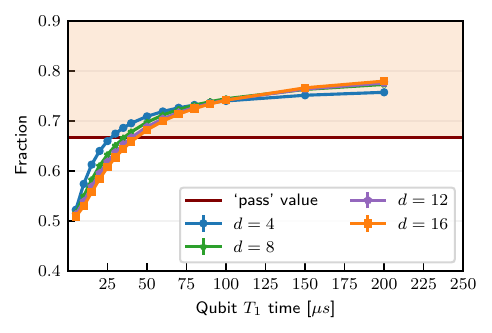}
\caption{Fraction of bitstrings lying in their corresponding heavy set for various T${}_1$ times and dimensions $d$. The bootstrapped error bars are negligibly small.}
\label{fig:hog_t1}
\end{figure}

It is evident that the metrics used to gauge the qudit's performance depend on the coherence of the ancilla qubit. The degree to which other parameters impact these metrics is often less pronounced.
The HOG metric is convenient in that the scheme provides a hard pass/fail value of two-thirds for the heavy output fraction\footnote{\label{fnt:hogtwothirds}This value arises from a combination of a worst-case approximation of the underlying binomial distribution as well as a chosen z-score of 2 or greater~\cite{cross2019validating}.}. Figure~\ref{fig:hog_t1} shows the results of the HOG test fraction for various targeted qudit dimensions $d$ as a function of the qubit's T${}_1$ time (where we have dropped the pure qubit dephasing contribution in Eq.~\eqref{eqn:lindblad}), averaged over an ensemble of $1000$ Haar random circuits for each data point.

First, it is unsurprising that the HOG fraction increases given less noisy qubits. This is a result of the phase kickback mechanism --- which is fundamental to the operation of the SNAP gate --- becoming more coherent for larger qubit T${}_1$ times and hence imparting the desired phases to the cavity mode Fock states. However, what is perhaps surprising is that relatively modest qubit T${}_1$ values on the order of $25-50 \ \mu$s --- values far from state-of-the-art --- appear sufficient to achieve relatively good control (``good control'' here gauged with respect to the HOG test pass/fail value) over the first $\sim 10$ Fock levels. At a T${}_1$ value of 100~$\mu$s, typical for modern transmons, all simulated dimensions exceed the ``pass'' value with a significant margin, pointing to the possibility of controlling $\sim 10^2$-dimensional qudits.

It is instructive to note the relationship between the expected controllable dimension of the cavity qudit, and the T${}_1$ time of the coupled transmon qubit. In our simulations, a $\chi$ of $2\pi \times 1 \ \text{MHz}$ was chosen. As such, a single SNAP pulse had a length of $10 \sqrt{d} / (1 \ \text{MHz}) = 10 \sqrt{d} \ \mu$s (recall that we scale the length of each SNAP pulse to control the total microwave power deposited into the system). In order to reduce incoherent errors due to the transmon's decay, we would like our SNAP gate time to be less than the decay time scale, i.e. $10 \sqrt{d} < \mathrm{T}_1$, or equivalently $d < (\mathrm{T}_1/10)^2$. In contrast, since T${}_1$ times are inversely proportional to the average photon number, or roughly the cavity qudit dimension, one may expect that the ratio of the cavity qudit T${}_1$  ($\sim 1$~s) to that of the transmon qubit ($\sim 100~\mu$s), which is roughly $10^4$, provides an estimate of the controllable dimension of the cavity qudit.

Given the relationship between the transmon T${}_1$ and the qudit dimension $d$ that we just discussed, however, achieving $d \sim 10^4$ translates to a transmon T${}_1 \sim 1$ ms. While this is larger than currently available T${}_1$ times, transmons optimized for coherence can regularly achieve T${}_1$ times in excess of 100 $\mu$s~\cite{carroll2022quantum, Houck2021transmon, Wang2022transmon, Bal2024systematic}. The current state-of-the-art is $> 500~\mu$s~\cite{Wang2022transmon, Bal2024systematic},  and therefore should be more than sufficient for a qudit dimension of $d \sim 100$. 
Promising directions to achieve even higher T${}_1$ times include side encapsulation of the base superconducting layer, substrate treatment to reduce bulk loss, modification of the device geometry, fabricating Josephson junctions with materials other than Aluminum etc. This suggests that bad qubit decay rates are potentially not the largest bottleneck when creating arbitrary Fock states of higher dimension $d$ in such a cavity-qubit system. 
However, it should be noted that many other minor sources of errors such as state readout error, non-zero thermal photon noise, control electronics imperfections etc. are not accounted for in this model and may mildly decrease the HOG metric.

As our numerical studies above suggest, increasing the target qudit dimension $d$, for a fixed T${}_1$, appears to have a relatively modest deleterious effect on the HOG fraction. However, it may not be the case in practice because increasing $d$ would entail increasing the number of Fock levels to which arbitrary SNAP phases need to be applied. The calibration of multiple simultaneous SNAP pulses is challenging in the laboratory due to the presence of AC-Stark shifts in the qubit's energy spectrum, which often gives rise to other unwanted effects~\cite{carroll2022dynamics}. Studying other potential gatesets such as Echoed Conditional Displacement~\cite{ECD2022NatPhys} operations along with single-qubit rotations, the description of which is more natural in terms of canonical coherent states rather than Fock states, would be interesting.

\begin{figure}[t]
\centering
\includegraphics[width=0.7\linewidth]{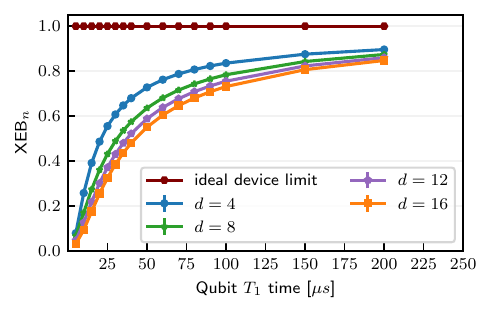}
\caption{Normalized linear cross-entropy as a function of qubit T${}_1$ time (with no qubit dephasing T${}_2$ channel), with small bootstrapped error bars.}
\label{fig:xeb_t1}
\end{figure}

Figure~\ref{fig:xeb_t1} plots the $\text{XEB}_n$ metric from the same datasets used in Fig.~\ref{fig:hog_t1}. This metric, too, improves with a more coherent qubit. Once again, a relatively modest T${}_1$ time appears capable of yielding appreciable linear cross-entropy values. However, for some fixed T${}_1$ value, one cannot arbitrarily keep increasing the targeted cavity qudit dimension without sacrificing controllability on its Hilbert space, as captured by the $\text{XEB}$ metric.

The error bars in Figs.~\ref{fig:hog_t1} and \ref{fig:xeb_t1} are derived from bootstrapping the $1000$ individual HOG fraction/$\text{XEB}_n$ values, for each data point. We do not report the standard deviation of these 1000 values since each one arises from an independent unitary with a different underlying distribution; we are interested in investigating the performance of the cavity-transmon device itself, rather than how well the device can prepare any particular state. As such, we employed a Bayesian bootstrapping procedure outlined in the Appendix Sec. \ref{secn:stats} to derive the bootstrapped standard deviation error bars shown in these two figures. These bars give an indication of how reliably the device itself performs and how well it is able to consistently produce the HOG/$\text{XEB}_n$ values for the various values of T${}_1$ and $d$. Given that these deviations are small (roughly $0.5-1\%$ for each data point in each of the two figures), we may be confident that we have sampled a sufficient number of Haar random unitaries and that the device's performance is relatively consistent.

\begin{figure}[t]
\centering
\includegraphics[width=0.7\linewidth]{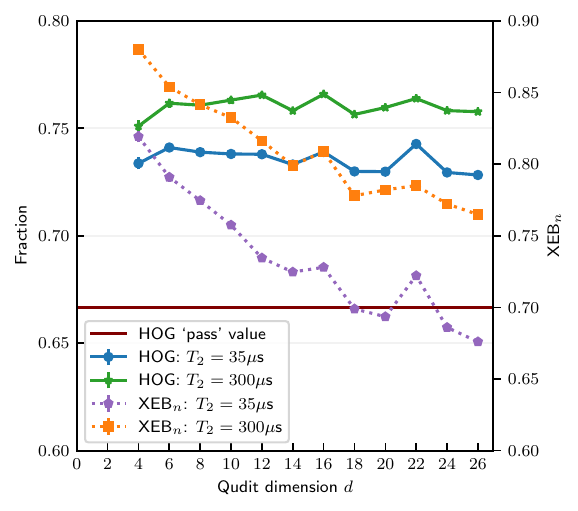}
\caption{Heavy output test fraction and normalized linear cross-entropy metric for a cavity-qubit system, all with T${}_1 = 150 \ \mu$s and T${}_2$ values as indicated.}
\label{fig:hog_and_xeb_realistic}
\end{figure}

Finally, we simulate the dynamics and investigate both metrics for a device with a modest qubit T${}_1$ value of $150 \ \mu$s~\cite{Bal2024systematic}, and two different qubit T${}_2$ values: one relatively poor value ($35 \ \mu$s) and the other ($300~\mu$s) chosen such that the pure dephasing time T${}_{\phi}$ \cite{krantz2019quantum} (where $\frac{1}{\mathrm{T}_{\phi}} = \frac{1}{\mathrm{T}_{2}} - \frac{1}{2\mathrm{T}_{1}}$), is zero. Figure~\ref{fig:hog_and_xeb_realistic} shows the metrics' behaviors as a function of $d$. Note that T${}_2$ values appear to have relatively small effects on the performance on the sampling tests we consider here. For instance, Fig.~\ref{fig:hog_and_xeb_realistic} suggests that the device passes the HOG test for the same $d$ even after an order of magnitude increase in the T${}_2$ value.

As in the previous figures, the error bars in Fig.~\ref{fig:hog_and_xeb_realistic}, obtained from bootstrapping, are negligibly small. Despite this, we observe non-negligible variations in the data at, for example, $d=16$ and $22$. We confirmed that this deviation is not due to compilation errors of the ideal Haar random target states, since each state was post-selected on having been compiled to the pulse level with an infidelity of less than 1\%, which we deemed acceptable. Furthermore, we ensured that this post-selection would not introduce any bias in the final results by applying a so-called $t$-design frame potential test. Unitary \(t\)-designs provide a systematic method to approximate Haar-random unitaries. In particular, given an ensemble of unitaries, the frame potential test helps to quantify how close this ensemble is to being truly Haar-random. These numerical tests (see Fig.~\ref{fig:frame-potential-test} in Appendix Sec.~\ref{app:haarrandomandt}) confirm that our postselection does not bias the uniformity of the ensemble of unitaries used in the HOG test. This test also confirmed that sampling 1000 states, for each data point, was sufficient.
Furthermore, such variations cannot be due to finite sampling errors arising from insufficient measurements of the final state's Fock level, given that we had full knowledge of the noisy density matrices themselves and computed the HOG and XEB metrics directly (although the Bayesian inference protocol one would need to implement in a real-world setting to infer the posterior distribution, is outlined in the Appendix Sec. \ref{secn:stats}, for completeness).

The overall trends in Fig.~\ref{fig:hog_and_xeb_realistic} show that, given a fixed T${}_1$ time, a stronger dephasing noise reduces the controllability of the qudit dimension. In contrast, in the case of no pure qubit dephasing, the benchmarks perform better for all target qudit dimensions. In summary, the three figures viewed together suggest that one should be able to control a few tens of Fock states of a high-quality cavity mode using a transmon with contemporary decoherence values, without regard to other possible noise channels.

\section{Conclusion and Outlook}
\label{secn:conclusion}

In this paper, we tested the ability of a qudit-based QPU, consisting of an ancilla transmon qubit dispersively coupled to a cavity resonator, to generate ensembles of random states in the Fock basis. These were in turn investigated using modified versions of popular sampling tests that are often used, in the current NISQ-era, to gauge QPU device performance. In particular, we adapted the heavy output generation test and the linear cross-entropy benchmark, previously used for qubit-based platforms, to this qudit-based one. These results were numerically studied using a full-stack of quantum processing elements at the pulse-level, under a realistic noise model. We found that under reasonable assumptions about the noise of the cavity-transmon system, one can manufacture a cavity qudit with reliable controllability of its Hilbert space containing a few tens of Fock levels with contemporary transmons. We expect this dimensionality to go even higher as coherence times improve on transmon qubits~\cite{Houck2021transmon, Wang2022transmon, Bal2024systematic}.

Our study used classical simulation of a quantum device under a simple, but realistic, noise model. We found intuitively plausible improvement of the metric values under reduced noise. Our noise model assumed a very high coherence cavity, with the primary limitations arising from noise on the transmon qubit. This arrangement is expected based on current hardware implementations of these devices~\cite{Reagor2016coax, Rosenblum2023Mushroom, Oriani2024Nb_cav, Yu2024stabilization, Roy2024Qudit}. The most important next step is to apply the framework developed here to physical hardware in the laboratory and assess how well one can control individual qudits with increasing dimension.

It is also important to consider the performance of the metrics in a system of entangled cavity resonators. As the system scales to multiple qudits~\cite{Reineri2023multimode, you2024crosstalk}, one could then employ random qudit circuits ~\cite{Harrow2023}, with Haar random unitaries on individual qudits, as we have done here, and entangling gates between the qudits, to perform quantum volume experiments analogous to the qubit case. Luckily, as the pathway to operable multi-mode systems becomes clearer, these same tools should be relatively easily generalizable, making possible comparisons between single- and multi-mode devices. While a specific qudit platform with a specific gateset and noise model was chosen, the underlying random sampling tests should still hold for different qudit platforms with different gatesets. This is an advantage of our random sampling, algorithm-agnostic benchmarking approach. Moreover, certain other traditional benchmarking techniques such as randomized benchmarking \cite{magesan2011scalable,magesan2012characterizing} face difficulties in the continuous variable setting \cite{iosue2024continuous}. These difficulties can be circumvented by the sampling tests we have adopted here, which are also often key to contemporary quantum advantage demonstrations \cite{QSupremacy,PhysRevLett.127.180501,ZHU2022240}. Such tests, as outlined in this manuscript, may make future quantum advantage comparisons easier to compare between different devices.

For the sampling tests employed here, an important ingredient is the classical circuit simulation compute ideal probability distributions. In principle one could adopt standard tensor network techniques to compute individual amplitudes or sampling probabilities. However, larger qudit dimensions would increase the computational runtime of such methods, if not necessarily their asymptotic complexity. It also leaves room for improvement on qubit-based simulation algorithms, that may exploit certain features of the gateset we employ here. 
While the HOG/XEB benchmarks generate random quantum circuits (from one- and two-qubit random gates) in the multi-qubit setting, here we directly sampled single-qudit Haar random unitaries and compiled them to the native gateset. 

A natural research direction for the future would be to investigate the efficient construction of unitary/state $t$-designs \cite{ambainis_quantum_2007,scott_optimizing_2008,dankert_2009_exact,roy_unitary_2009,roberts_chaos_2017,mele_introduction_2023} that can approximate Haar random ensembles on single qudit Hilbert spaces in terms of SNAP and displacement gates, the native gateset employed here (see Appendix Sec.~\ref{app:haarrandomandt} for details). The optimal construction of qudit unitary designs has attracted a lot of attention recently, see e.g., the seminal work of Ref. \cite{Brando2016} and the more recent \cite{Harrow2023} including references therein. These are primarily based on ``brickwork" circuits generated via 2-qubit Haar random unitaries and it would be interesting to construct ``hardware efficient" variants that can generate them on cavity qudits via native interactions. It is worth mentioning that the infinite-dimensional single cavity mode was truncated and studied in the discrete photon-number basis in this study. In reality, the cavity mode is a continuous bosonic system. However, constructing continuous-variable (CV) designs for such systems is a nontrivial task and requires exquisite control over the measurements on a cavity mode \cite{Iosue2024}. While randomized benchmarking will likely remain an important tool, the sampling approaches we considered here offer a useful and readily accessible alternative for benchmarking hybrid cavity-transmon systems when viewed from a continuous-variable lens.

We also remark that while a circuit ansatz consisting of SNAP and displacement gates of depth $O(d^2)$ is expressive enough to compile arbitrary single qudit unitaries of dimension $d$, adopting a lower depth $O(d)$ depth ansatz would be less noisy while still maintaining expressivity. However, this comes at a cost of incurring compilation errors in practice, as there is no known classical algorithm that can efficiently find the globally optimal parameters for such an ansatz, unlike for the $O(d^2)$ ansatz. A crucial input to the scalability of such systems would be to develop compilation algorithms for such gatesets. We leave all such considerations for future work.

\section*{Acknowledgments}
\label{sec:ack}

We thank Erik Gustafson and Henry Lamm for their helpful feedback, and Taeyoon Kim for his assistance with the pulse-level compilation.

This material is based upon work supported by the U.S. Department of Energy, Office of Science, National Quantum Information Science Research Centers, Superconducting Quantum Materials and Systems (SQMS) Center under the contract No. DE-AC02-07CH11359.  This document was prepared using the resources of the Fermi National Accelerator Laboratory (Fermilab), a U.S. Department of Energy (DOE), Office of Science, HEP User Facility. Fermilab is managed by Fermi Forward Discovery Group,1039
LLC under Contract No. 89243024CSC000002 with the U.S. Department of Energy, Office of Science, Office of High Energy Physics. N.A. is a KBR employee working under the Prime Contract No. 80ARC020D0010 with the NASA Ames Research Center. M.S.A. acknowledges support from USRA NASA Academic Mission Services under contract No. NNA16BD14C with NASA, with this work funded under the NASA-DOE inter-agency agreement SAA2-403602 governing NASA’s work as part of the SQMS Center.

\bibliographystyle{iopart-num}
\bibliography{refs}


\appendix

\section{Bayesian inference and bootstrapping setup}
\label{secn:stats}

Given that we simulate the QPU classically and hence know the final noisy state $\rho_U$ exactly for the cavity qudit for each unitary $U$, as well as the classical Fock state distribution $p_U$ derived from it, there is no need in the current article to consider finite sampling statistics. However, one needs to employ statistical techniques to infer $p_U$ when performing such experiments in practice.
In that case we won't have direct access to $\rho_U$ but may only sample from $p_U$ repeatedly, in a particular measurement basis (here, the photon number basis). We here give a pedagogical outline of the Bayesian inference techniques needed to glean $p_U$ for each unitary $U$. Furthermore, we outline a Bayesian bootstrapping procedure \cite{rubin1981bayesian} to assign error metrics for finite sampling of unitaries $U$ from the QPU itself.

For a given circuit described by the $d \times d$ unitary $U$, acting on a qudit initialized in the ground state $\ket{0}$, the ideal noiseless bitstring probability vector $\vec{q}_U$ is given by Eq.~\ref{eqn:idealbitstringprob}. The probability of an output bitstring $i \in \{ 0, 1, \cdots, d-1 \}$ in the noisy implementation of this circuit follows a different distribution, denoted by vector $\vec{p}_U$ with elements $p_{U,i}$. The act of sampling a single discrete bitstring $x_U$ from one of the $d$ categories, each with corresponding probability $p_{U,i}$, is modeled as sampling from a categorical distribution, i.e. $x_U \sim \textrm{Categorical}(\vec{p}_U)$; drawing $N$ independent and identically distributed bitstrings $\vec{x}_U$ corresponds with sampling from the corresponding multinomial distribution, $\vec{x}_U \sim \textrm{MultiNom}(N, \vec{p}_U)$.

The prior distribution most appropriate for modeling the uncertainty of $\vec{p}_U$ itself, given that we sample from a multinomial distribution in the lab, is a Dirichlet distribution as it is the conjugate prior distribution. So, $\vec{p}_U \sim \textrm{Dir}(\alpha_0, \cdots, \alpha_{d-1})$ with the set of $\alpha_i$'s being the concentration hyperparameters. 
As such, assuming the prior distribution $\textrm{Dir}(\alpha_0, \alpha_1, \cdots)$ and having observed the sample vector $\vec{x}_U$ consisting of the number of recorded observations $x_{U,i}$ from each of the $d$ categories (with $\sum_{i=0}^{d-1}x_{U,i} = N$), the updated posterior distribution is $\textrm{Dir}(\alpha_0 + x_{U,0}, \alpha_1 + x_{U,1}, \cdots)$. The choice of $\vec{\alpha}$ is up to the user, generally.

One common choice is to set $\alpha_i=0$ for all $i$. This is an extension of the Haldane prior, which is an improper distribution often used in binomial sampling (i.e., repeatedly sampling a binary variable). This prior has the property that the expectation value of the posterior distribution for each outcome exactly matches the sample mean. Since one cannot sample from an improper distribution, if we were to choose $\alpha_i=0$ and observe $x_{U,j}=0$ for some $j$, we'd have to ``drop'' such an element from the distribution. In other words, we'd sample over only those elements for which $x_{U,i}\neq 0$. Doing so would be equivalent to the Bayesian bootstrap from \cite{rubin1981bayesian} whereby one sampled an indistinguishable element repeatedly.

Another hyperparameter choice is to set $\alpha_i = 1/d$ for the $d$ categories. This is a common choice for a prior such that the resulting Dirichlet distribution matches the marginal distributions for each category \cite{gelmanbayesian2013}. Given that choosing a prior's hyperparameters is largely up to the user, and for reasons of simplicity, this is our choice.

The ``target'' distribution $\vec{q}_U$, as well as the above posterior, are enough to compute the probability inner product from Sec. \ref{subsecn:xeb} for a single $U$, according to

\begin{equation}
\langle p_U, q_U \rangle \equiv p_U(x) \cdot q_U(x) = \int \left(\vec{q}_U \cdot \vec{p}_U \right) f(\vec{p}_U; \vec{\alpha} + \vec{x}_U) d\vec{p}_U,
\end{equation}

\noindent
where the probability density function $f$ of the posterior is given by (with $A = \sum \alpha_i$)

\begin{equation}
f(p_{U,0}, \cdots, p_{U,d - 1}; \alpha_0, \cdots, \alpha_{d - 1}) = \frac{\Gamma(A)}{\prod_{i=0}^{d-1}\Gamma(\alpha_i)} \times \prod_{i=0}^{d - 1} p_{U,i}^{\alpha_i - 1}.
\end{equation}

However, in the case of the HOG test (see Sec. \ref{subsecn:hog}), further simplifications can be made. Here, we first classically compute the heavy set $H_U$ (Eq.~\ref{eqn:hogheavyset}). We are interested in what fraction of the bitstrings that we sample lie in $H_U$. Luckily, the marginal distribution for a bitstring being in $H_U$, given our posterior distribution $\textrm{Dir}(\vec{\alpha}+\vec{x}_U)$, is simple to describe. Indeed, for a general Dirichlet distribution over distinct categories $\{A, B, C, \cdots\}$, namely $\textrm{Dir}(a,b,c,\cdots)$, the marginal distribution $\textrm{Pr}(A \textrm{ OR } B, C, D, \cdots)$ formed by identifying categories $A$ and $B$, is simply $\textrm{Dir}(a+b,c,\cdots)$. In our case, we simply aggregate all bitstrings into two categories: those that lie in $H_U$ and those that do not. The marginal distribution for a bitstring being in $H_U$ is hence simply $\textrm{Dir}(\alpha + \sum_{i \in H_U} x_{U,i}, \alpha + \sum_{i \notin H_U} x_{U,i})$. Or equivalently\footnote{\label{fnt:bayesian}The Dirichlet distribution reduces to the Beta distribution for two categories: the Dirichlet distribution Dir$(x_1,x_2)$ samples $p_1$ and $p_2$, but $p_1+p_2=1$. Therefore one needs only model a distribution for $p_2$, which is the Beta distribution Beta$(x_1,x_2)$}, the fraction of measurements in the heavy output has posterior $p_{U,\text{heavy}} \sim \textrm{Beta}(\alpha + \sum_{i \in H_U} x_{U,i}, \alpha + \sum_{i \notin H_U} x_{U,i})$ (with $\alpha = 1/2$ for this two category case).

Recall that we did not have to consider the sample size $N$ in the main text as our experiments were numerically simulated and we hence had access to the exact noisy distribution $\vec{p}_U$. However, in practice, how many bitstring measurements, $N$, would we need to take from each noisy implementation of $U$ with the posterior $\vec{p}_U \sim \textrm{Dir}(1/d + x_{U,0}, 1/d + x_{U,1}, \cdots)$ in order to bound the error of each bitstring probability $p_{U,i}$? In this case, $p_{U,i}$'s expectation value and variance are given by

\begin{align}
\text{E}\left[ p_{U,i} \right] & = \frac{\frac{1}{d} + x_{U,i}}{N+1}, \\
\text{Var} \left[ p_{U,i} \right] & = \frac{(N+1)(\frac{1}{d} + x_{U,i}) - (\frac{1}{d} + x_{U,i})^2}{(N+1)^2(N+2)}.
\end{align}

This variance attains its maximum when $x_{U,i}$ is an integer close to $\frac{1}{2}(N+1) - \frac{1}{d}$. Employing Chebyshev's inequality, we find

\begin{equation}
\text{P}(\left| p_{U,i} - \text{E}\left[ p_{U,i} \right] \right| \geq \epsilon) \leq \frac{\text{Var} \left[ p_{U,i} \right]}{\epsilon^2} \leq \frac{1}{\epsilon^2}\frac{1}{4(N+2)}.
\end{equation}

\noindent
For example, suppose that we wish the estimate of $p_{U,i}$ to be within $0.1$ of the true value with $99\%$ certainty. In this case it is sufficient to take $N \sim 2,500$.

Finally, each circuit $U$ gives rise to a different $\vec{q}_U$ and $\vec{p}_U$. We therefore need to repeat the above analysis for each circuit. With this in mind, one question remains: how do we combine the individual HOG fractions or XEB ``inner products'' for distinct $U$'s to arrive at the QPU's overall metric? One approach is to simply find the sample mean of the individual HOG fraction or inner product values. However, this sample mean is over an ensemble of HOG fractions and inner product values which all come from different probability distributions. We wish to make inferences about the QPU itself which with some unknown fixed distribution has given rise to the different circuits, both ideal and noisy. Since we know nothing about the distribution across the different circuits a priori, we will use a Bayesian bootstrapping technique over the circuits. 

In general, bootstrapping involves random resampling, with replacement, of the original sample in order to compute various accuracy metrics of the sample. This takes place with the fundamental tenet of statistical bootstrapping in mind: any bootstrapped statistic is to the original sample statistic as the original sample statistic is to the population's statistic. This is only accurate if we take the sample of $n$ random unitaries/circuits to be a reasonable approximation of the QPU's unknown population distribution. 

Rubin's Bayesian bootstrap has a similar flavor \cite{rubin1981bayesian}. It takes the observed elements - here the $n$ random unitaries, or values/metrics derived from them - as a given (i.e., the distribution only has support on the observed elements). However, since we observed each element only once (assuming they're all distinguishable), we have learned nothing about their relative frequency in the true distribution. Thus, Rubin suggests performing a bootstrap by re-weighting each sample value according to a Categorical probability distribution sampled uniformly over the simplex.

In our case the posterior estimate of the QPU's heavy output probability, $p_{\text{heavy}}$, over all circuits, is $p_{\text{heavy}} \sim \sum_{i=1}^{n} \omega_{i} \, p_{U_{i},\text{heavy}}$, where the $n$-dimensional weight vector $\vec{\omega}$ is sampled from $\text{Dir}(1, 1, \cdots, 1)$, a posterior that is flat over the $n-1$ simplex.  As mentioned above, this is equivalent to using an improper extended-Haldane prior over the space of outcomes and dropping those elements that are not observed, in order to render the posterior a proper distribution. The posterior for the QPU's overall linear cross-entropy metric is similar: $\text{XEB} \sim \sum_{i=1}^{n} \omega_{i} \left(d \langle p_U, q_U \rangle - 1 \right)$. These distributions can be used to assign error bars to the QPU's heavy output and cross-entropy metrics. The interested reader is directed to \cite{gelmanbayesian2013} for more details on Bayesian techniques.

\section{Haar random unitaries and \texorpdfstring{$t$}{t}-designs}
\label{app:haarrandomandt}

\begin{figure}[t]
\begin{subfigure}{.5\textwidth}
\centering
\caption{Qudit dimension \(d\) = 4}
\includegraphics[width=\linewidth]{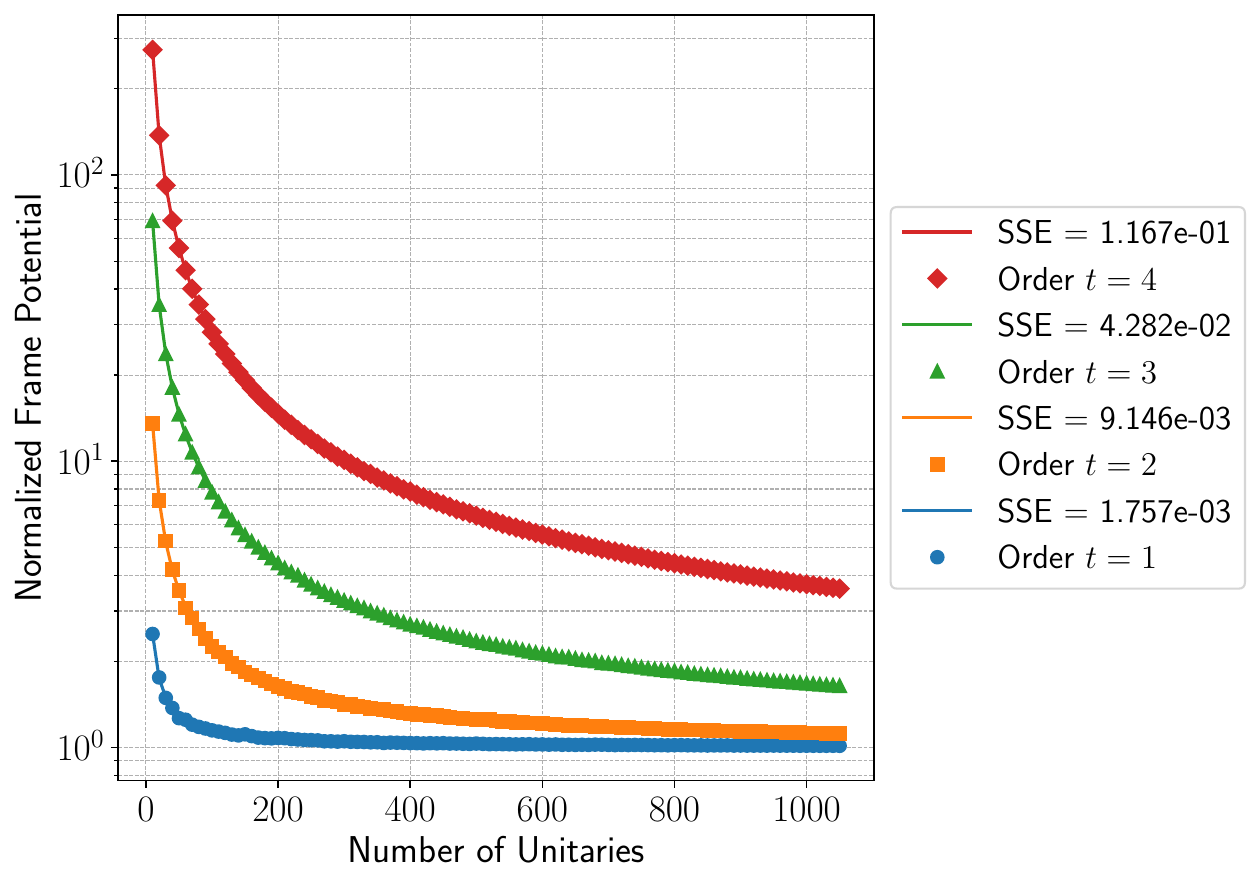}  
\label{fig:qudit-dim-4-tdesign}
\end{subfigure}
\begin{subfigure}{.5\textwidth}
\centering
\caption{Qudit dimension \(d\) = 8}
\includegraphics[width=\linewidth]{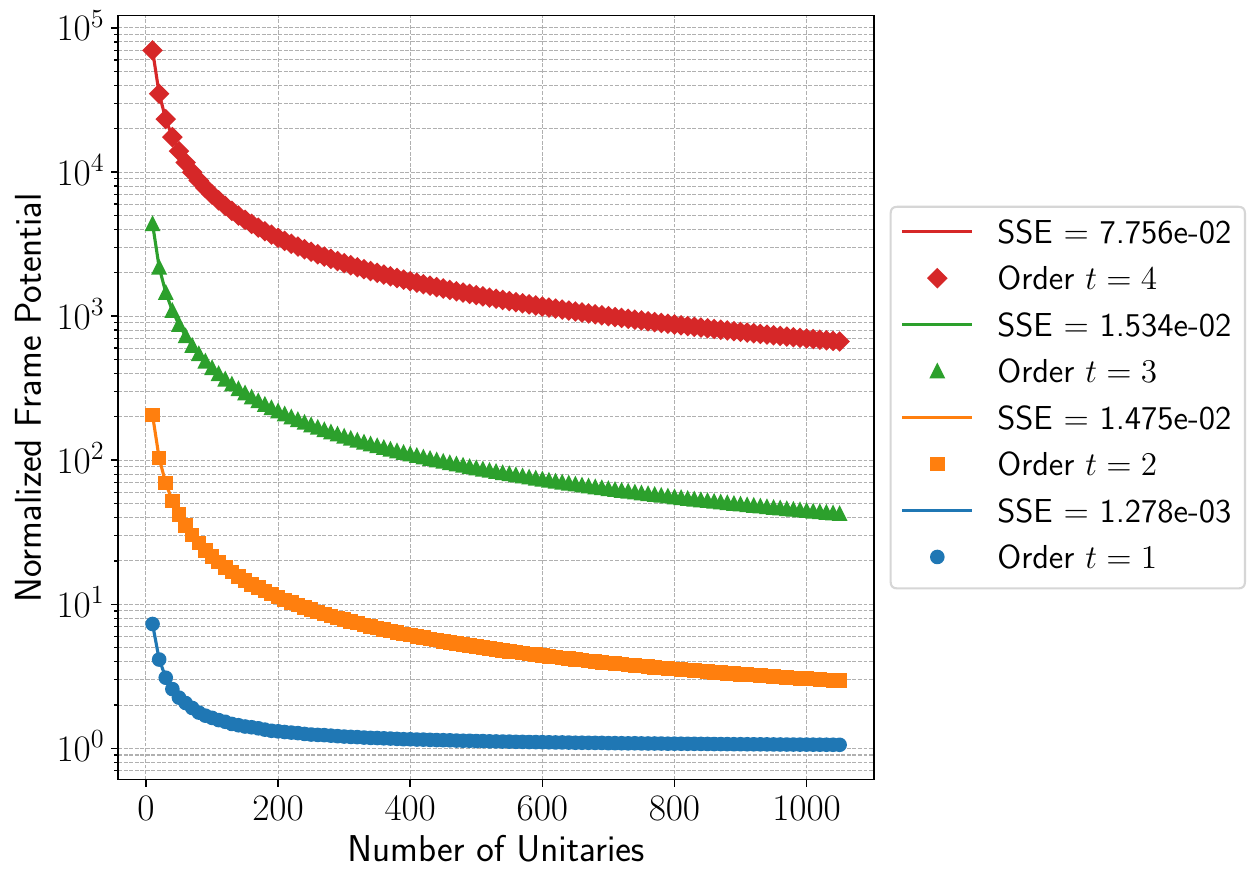}  
\label{fig:qudit-dim-8-tdesign}
\end{subfigure}
\begin{subfigure}{.5\textwidth}
\centering
\caption{Qudit dimension \(d\) = 12}
\includegraphics[width=\linewidth]{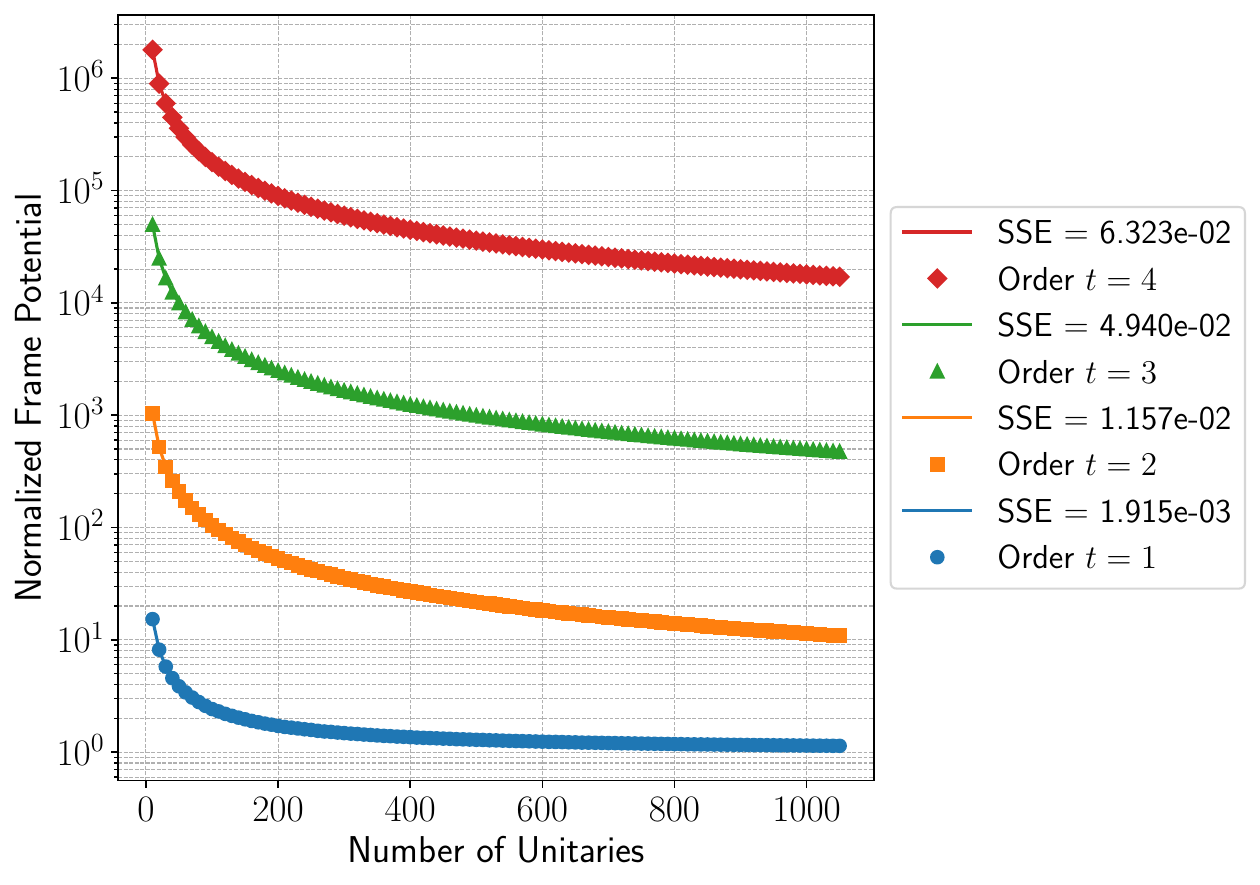}  
\label{fig:qudit-dim-12-tdesign}
\end{subfigure}
\begin{subfigure}{.5\textwidth}
\centering
\caption{Qudit dimension \(d\) = 16}
\includegraphics[width=\linewidth]{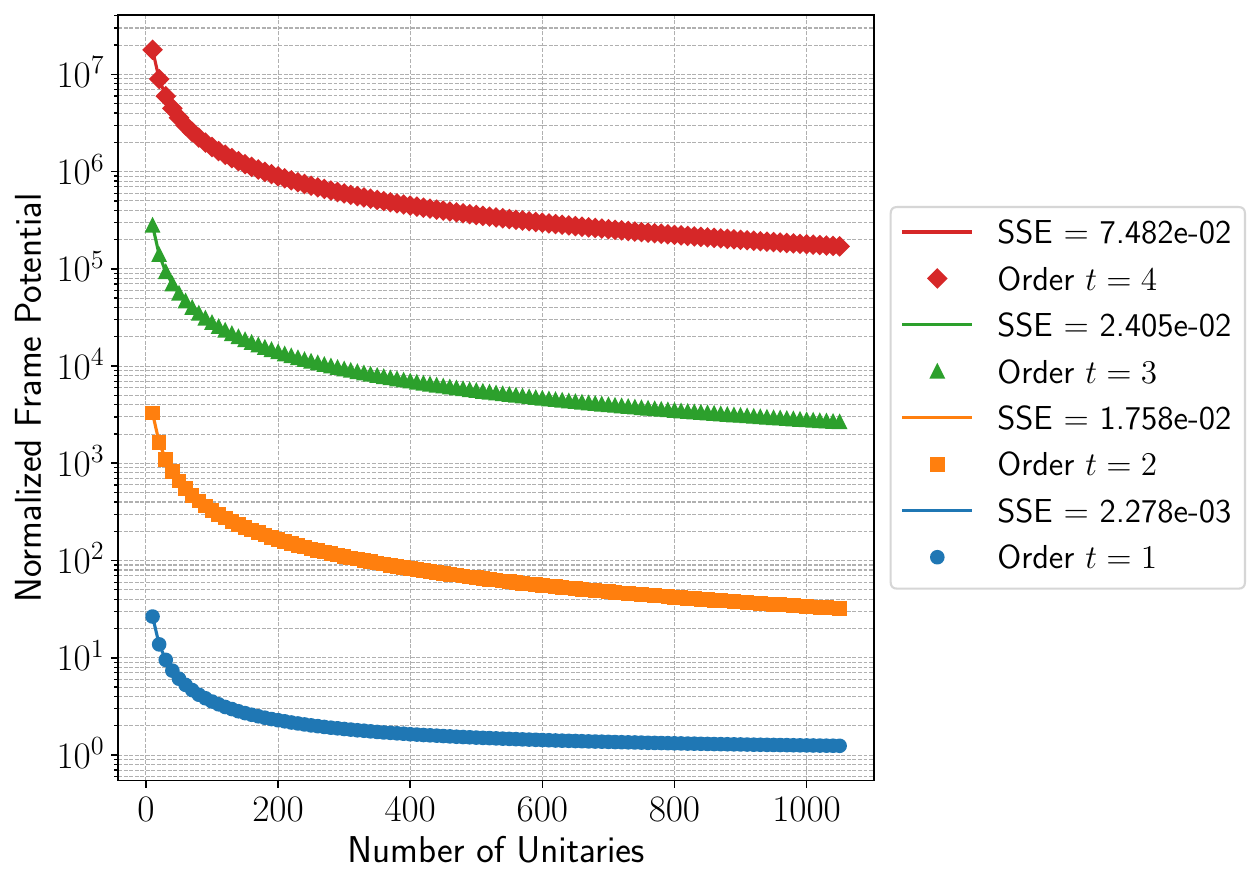}  
\label{fig:qudit-dim-16-tdesign}
\end{subfigure}
\caption{Normalized frame potential of the ensemble of ``filtered'' unitaries that are used in our HOG test. We plot the normalized frame potential, \(\mathcal{F}^{(t)}_{\mathcal{E}}/\mathcal{F}_{\mathrm{Haar}}^{(t)}\) for \(t \in \{1,2,3,4\}\) and for qudit dimensions \(d=\{4,8,12,16\}\). We sample all of the \(1050\) unitaries generated for the HOG test. All the curves asymptotically converge to one, which signifies that they behave as Haar-random unitaries. We fit these decay curves to a functional form \(\mathcal{F}^{(t)}_{\mathcal{E}}(N)=1+\frac{A}{N^{p}}\), where \(N\) is the number of samples. The legends also list the sum of squared errors (SSE) for each of the fits. The fit parameters are close to the values obtained by sampling the same number of elements from the Haar-random ensemble. This confirms that our filtering does not bias the uniformity of the random unitaries used in the HOG test.}

\label{fig:frame-potential-test}

\end{figure}

The Haar measure on the unitary group \(U(d)\) is the unique (normalized) group invariant measure that allows us to sample unitaries uniformly over a \(d\)-dimensional Hilbert space, \(\mathcal{H} \cong \mathbb{C}^{d}\). Unitaries drawn from this distribution are commonly referred to as ``Haar random unitaries''. Similarly, applying a Haar random unitary \(V\) to a fixed reference state \(| \psi_{0} \rangle\) generates a state \(| \phi \rangle = V | \psi_{0} \rangle\) which is also uniformly distributed (over \(\mathcal{H}\)). Since the Haar measure is unitarily invariant, this means that the states \(| \phi \rangle\) obtained in this manner are also unitarily invariant and independent from the reference state \(| \psi_{0} \rangle\). Such states are colloquially known as ``Haar random states''. They have a number of unusual properties such as being nearly maximally entangled across any bipartition \cite{Hayden2006}, having nearly maximal magic (or nonstabilizerness) \cite{Leone2022}, and having close to maximal quantum coherence \cite{Singh2016}, among others. 

These unusual properties make Haar random states and unitaries a common tool in quantum information theory, see e.g., the excellent review \cite{mele_introduction_2023}. Unfortunately, these states and unitaries suffer from a crucial drawback: their quantum circuit complexity grows \textit{exponentially}. Therefore generating them even on a quantum computer is not an efficient task. Although this is not an issue for the (low-dimensional) single qudit systems considered in this work, this can be an issue when generalizing to multi-qudit and large dimensional systems.
This bottleneck is at least partly resolved with the invention of the so-called ``unitary \(t\)-designs'' which approximate Haar random unitaries up to the \(t\)-th moment, where \(t \in \mathbb{N}\) \cite{ambainis_quantum_2007,scott_optimizing_2008,dankert_2009_exact,roy_unitary_2009,roberts_chaos_2017,mele_introduction_2023}. 
Therefore, if one is only interested in low-degree polynomial properties arising from Haar random states/unitaries, they can be efficiently generated quantumly. For multi-qubit systems some canonical examples are: the Pauli group forms a \(1\)-design and the Clifford group forms a \(2\)- and \(3\)-design. Moreover, by using random circuit constructions, one can generate polynomial orders of \(t\)-designs. This also provides the theoretical underpinnings of quantum supremacy and random circuit sampling experiments \cite{Brando2016,QSupremacy}.

This brings us naturally to the question of quantifying how close an ensemble of unitaries \(\mathcal{E}\) is to being Haar randomly distributed. A standard way to quantify this is to measure the $t^{\text{th}}$ order ``frame potential'' of the ensemble, namely,

\begin{align}
\mathcal{F}_{\mathcal{E}}^{(t)}:=\frac{1}{|\mathcal{E}|^{2}} \sum_{U, V \in \mathcal{E}}\left|\operatorname{Tr}\left[U^{\dagger} V\right]\right|^{2 t}, \quad t \in \mathbb{N},
\end{align}

\noindent
where $|\mathcal{E}|$ is the cardinality of $\mathcal{E}$.
The frame potential for Haar random unitaries has a closed form expression as, \(\mathcal{F}_{\mathrm{Haar}}^{(t)} = t!\) for \(t \leq d\). Moreover, we have, \(\mathcal{F}^{(t)}_{\mathcal{E}} \geq \mathcal{F}_{\mathrm{Haar}}^{(t)} ~~\forall t, ~~\forall \mathcal{E}\) with equality if and only if the ensemble is a unitary \(t\)-design.
One may then consider the ratio \(\mathcal{F}^{(t)}_{\mathcal{E}}/\mathcal{F}_{\mathrm{Haar}}^{(t)}\) to quantify the distance from the Haar ensemble. At this point it is worth mentioning that numerically estimating these quantities requires a number of samples that typically scales with the dimension of the Hilbert space and a finite but large number of samples may generate values away from the formulae listed above, see e.g., the numerical work and techniques developed in Ref. \cite{Liu2022}. As a result, one often tests for the asymptotic convergence to the Haar-random value, namely, \(\lim_{N \rightarrow \infty}\mathcal{F}^{(t)}_{\mathcal{E}}(N) \rightarrow \mathcal{F}_{\mathrm{Haar}}^{(t)}\) instead of the actual finite-sample value of the frame potential \(\mathcal{F}^{(t)}_{\mathcal{E}}(N)\). To check this numerically we fit the frame potential date to a functional form \(\mathcal{F}^{(t)}_{\mathcal{E}}(N)=1+\frac{A}{N^{p}}\), where \(N\) is the number of samples. We find that each of the curves in Fig. \ref{fig:frame-potential-test}, corresponding to various values of the frame potential \(t\) and qudit dimension \(d\), all converge to the Haar-random value asymptotically. This confirms that our numerical filtering does not bias the unitaries away from being uniformly distributed.

\section{Perfect compilation of SU(d) unitaries into SNAPs and SO(2)s}

Here, we review how to compile arbitrary unitaries using SNAP gates and $SO(2)$ rotations, often termed Givens rotations, as described in \cite{brennen2005criteria,spengler2010composite, SNAP2015PRA}. In principle, one could employ such a universal gateset and repeat the benchmarks described in the main text. However, the $SO(2)$ gates are not native to the hardware we consider, and would further need to be compiled into SNAP and Displacement gates \cite{SNAP2015PRA,job2023efficient}. On the other hand, as we show in this section, the advantage of using such a gateset would be that it would essentially eliminate all compilation errors, with the trade off of a larger circuit depth. We leave the study of such a gateset in realistic noisy conditions for future work, and describe the compilation procedure below.

The SNAP (Selective Number-Dependent Arbitrary Phase) gate \cite{SNAP2015PRL} is a diagonal gate that has the following representation
\begin{equation}
    S(\vec{\theta}) = \sum_{n=0}^{\infty} {e^{i\theta_n}} \vert n \rangle \langle n \vert,
\label{eqn:snap-gate}
\end{equation}
where $\vec{\theta} = (\theta_0, \theta_1, \dots, )$. In practice, of course, we would have $\theta_n = 0$ for all $n \geq d$ for some $d$. Typically, $d$ would be the dimensionality of the qudit space we wish to manipulate. As is clear from Eq.~\eqref{eqn:snap-gate}, the SNAP gate imparts an arbitrary phase $e^{i\theta_n}$ to each of the basis state vectors $\{ \vert n \rangle \}$. We will describe how this is achieved in the next section.

The second gate we wish to apply are $SO(2)$ rotations that are only active in the $\{\vert n \rangle, \vert n+1 \rangle \}$ subspace. These take the form
\begin{eqnarray}
    G_n(\alpha) &=& \mathbb{I}_{n} \oplus \begin{pmatrix}
    \cos{\alpha} & -\sin{\alpha}\\
    \sin{\alpha} & \cos{\alpha}
    \end{pmatrix} \oplus \mathbb{I}_{d-(n+2)},
\end{eqnarray}
where $\mathbb{I}_k$
is the $k \times k$ identity matrix, $0 \leq n \leq d-2$ and $d$ is the qudit dimensionality.

If our primitive gateset consists of $G_n(\alpha)$ for all $0 \leq n \leq d-2$ and arbitrary $\alpha$, and $S(\vec{\theta})$ for arbitrary $\vec{\theta} = (\theta_0, \dots, \theta_{d-1}, 0, 0, \dots)$, then we can compile any arbitrary $d \times d$ unitary matrix $U(d)$ into a product of these simpler operations exactly.
To demonstrate this, we first show how to compile $U^{\dagger}$. Consider any unitary
\begin{equation}
    U = \begin{pmatrix}
    a_{0,0}e^{i\theta_{0,0}} & a_{0,1}e^{i\theta_{0,1}} & \dots & a_{0,d-1}e^{i\theta_{0,d-1}} \\
    \vdots & & \ddots & \vdots \\
    a_{d-1,0}e^{i\theta_{d-1,0}} & a_{d-1,1}e^{i\theta_{d-1,1}} & \dots & a_{d-1,d-1}e^{i\theta_{d-1,d-1}}
    \end{pmatrix},
\label{eqn:snap-times-U}
\end{equation}
subject to the constraint $UU^{\dagger} = U^{\dagger}U = \mathbb{I}$. Let $\vec{\theta}_{k} \equiv (-\theta_{0,k}, -\theta_{1,k}, \dots, -\theta_{k,k})$ for $0 \leq k \leq d-1$. Then, we can ``de-phase" the entire last column of $U$ by left-multiplying it with a SNAP gate with parameter vector $\vec{\theta}_{d-1}$ so that
\begin{equation}
    S(\vec{\theta}_{d-1})\cdot U = \begin{pmatrix}
    a_{0,0}e^{i(\theta_{0,0}-\theta_{0,d-1})} & a_{0,1}e^{i(\theta_{0,1}-\theta_{0,d-1})} & \dots & a_{0,d-1} \\
    \vdots & & \ddots & \vdots \\
    a_{d-1,0}e^{i(\theta_{d-1,0}-\theta_{d-1,d-1})} & a_{d-1,1}e^{i(\theta_{d-1,1}-\theta_{d-1,d-1})} & \dots & a_{d-1,d-1}
    \end{pmatrix}
\end{equation}
Let us now note that we can transform an arbitrary 2-dimensional real vector to one whose top component is 0 and the bottom component is 1 using an $SO(2)$ rotation, i.e.
\begin{equation}
    \begin{pmatrix}
    \cos{\alpha} & -\sin{\alpha}\\
    \sin{\alpha} & \cos{\alpha}
    \end{pmatrix} \begin{pmatrix}
    a \\
    b
    \end{pmatrix} = \begin{pmatrix}
    0 \\
    1
    \end{pmatrix}
\end{equation}
for
\begin{equation}
\alpha = \tan^{-1}{\left( \frac{a}{b} \right)} = \cos^{-1}{\left( \frac{b}{a^2 + b^2}\right)} = \sin^{-1}{\left( \frac{a}{a^2 + b^2} \right)}
\end{equation}
Therefore, if we left-multiply the product in Eq.~\eqref{eqn:snap-times-U} by $G_{0}(\tan^{-1}{(a_{0,d-1}/a_{1,d-1})})$, then we transform $a_{0,d-1}\rightarrow 0$, $a_{1,d-1} \rightarrow 1$ and all other matrix entries to some other values. We can then repeat this process with $G_{1}, \dots, G_{d-2}$ with the angles for $G_{i}$ computed each time from the previous product of matrices $G_{i-1}(\alpha_{i-1}) \dots G_{0}(\alpha_0) S(\vec{\theta}_{d-1})$. 
When finished, we would be left with 0 in all the entries along the last column except for a 1 in the bottom-right corner of the matrix product. Since the entire product is unitary, this must also mean that the bottom-most row has all entries equal to 0 except (again) the rightmost element, which is equal to 1. At this point, we have
\begin{equation}
\prod_{i=d-2}^{0} G_{i}(\alpha_{i}^{(d-1)}) S(\vec{\theta}_{d-1}) U_{d} \equiv G_{d-2}(\alpha_{d-2}^{(d-1)}) \dots G_{0}(\alpha_{0}^{(d-1)}) S(\vec{\theta}_{d}) U_{d} = U_{d-1} \oplus I,
\end{equation}
where the superscript $(d-1)$ refers to the fact that are computing the $\alpha$ parameters for the $d-1$-th column. It is important to note that $\alpha_{i}^{(d-1)}$ cannot be computed until we have computed all the $\alpha_{j}^{(d-1)}$ for $j < i$ and composed the product of the corresponding unitaries with $S(\vec{\theta}_{d-1})\cdot U$. We can now repeat the entire process as above for $U_{d-1}$ leaving us with $U_{d-2} \oplus \mathbb{I}_{2}$, and then for $U_{d-2}$ and so on until we are just left with $\mathbb{I}_{d}$. 

At this point, we have compiled $U^{\dagger}$ as
\begin{equation}
U^{\dagger} = \prod_{j=d-1}^{0} \left( \prod_{i=j-1}^{0} G_{i}(\alpha_{i}^{(j)})\right) S(\vec{\theta}_{j}),
\end{equation}
and so therefore
\begin{equation}
U = \prod_{j=0}^{d-1} S(-\vec{\theta}_j) \left( \prod_{i=0}^{j-1} G_{i}(-\alpha_{i}^{(j)}) \right),
\label{eqn:snap-so2-compile}
\end{equation}
where $G_{i}(-\alpha_{i}^{(0)}) = \mathbb{I}$. We can readily observe from Eq.~\eqref{eqn:snap-so2-compile} that in order to compile an arbitrary $d \times d$ unitary, we need $d$ SNAP gates, and $\sum_{j=0}^{d-1} j = d(d-1)/2$ many $SO(2)$ gates. Each SNAP gate consisting of $k\leq d$ many non-trivial parameters may also require up to $2k$ many physical operations (see the section below). Thus, the physical gate cost of the SNAP operation may also scale as $d(d-1)/2$. By shaping the pulses required to realize this gate in a clever way, we can actually bring the physical gate cost down to just $2$, however. In any case, the total quantum gate cost is $O(d^2)$.

Although one may be able to apply the SNAP gate in $O(1)$ time, it may still require $O(d)$ to (classically) compute the parameters $\vec{\theta}_d$. Since $\vec{\theta}_{d-1}$ contains $d$ many non-trivial parameters, we have a total of $\sum_{j=1}^{d} = d(d+1)/2$ many classical computations to carry out for the SNAP gate. In addition, we also need $d(d-1)/2$ many classical computations to determine the $SO(2)$ rotation angles, one for each gate. Thus, the total classical computation cost of this compilation procedure is also $O(d^2)$.

It was recently shown by one of the authors that there is an efficient scheme to compile an arbitrary $SO(2)$ rotation into SNAPs and displacement gates without numerical optimization \cite{job2023efficient}. In essence, one can break the rotation $G_{k}(\theta)$ between states $\ket{k},\ket{k+1}$ into a repeated sequence of SNAPs and displacements of the form $V_{k}(\alpha)=D(\alpha)R_{\pi}(k)D(-2\alpha)R_{\pi}(k)D(\alpha)$, where $R_\pi(k)$ applies a SNAP with parameters $e^{i\pi}$ for all states up to and including $\ket{k}$. By setting $\alpha = \theta/4\sqrt{k+1}$, $V_k(\alpha)$ approximates $G_k(\theta)$ with an error that goes as $\theta^6$. Thus, breaking a larger rotation up into $m$ applications of $V_k(\theta/m)$. This compilation is constant time classically, and the total number of SNAPs and displacements for a constant error as a function of dimension $d$ is $\order{d^{2.5}}$.

\section{Truncating the displacement Gate}
\begin{figure}[!t]
\centering
\includegraphics[width=0.5\linewidth]{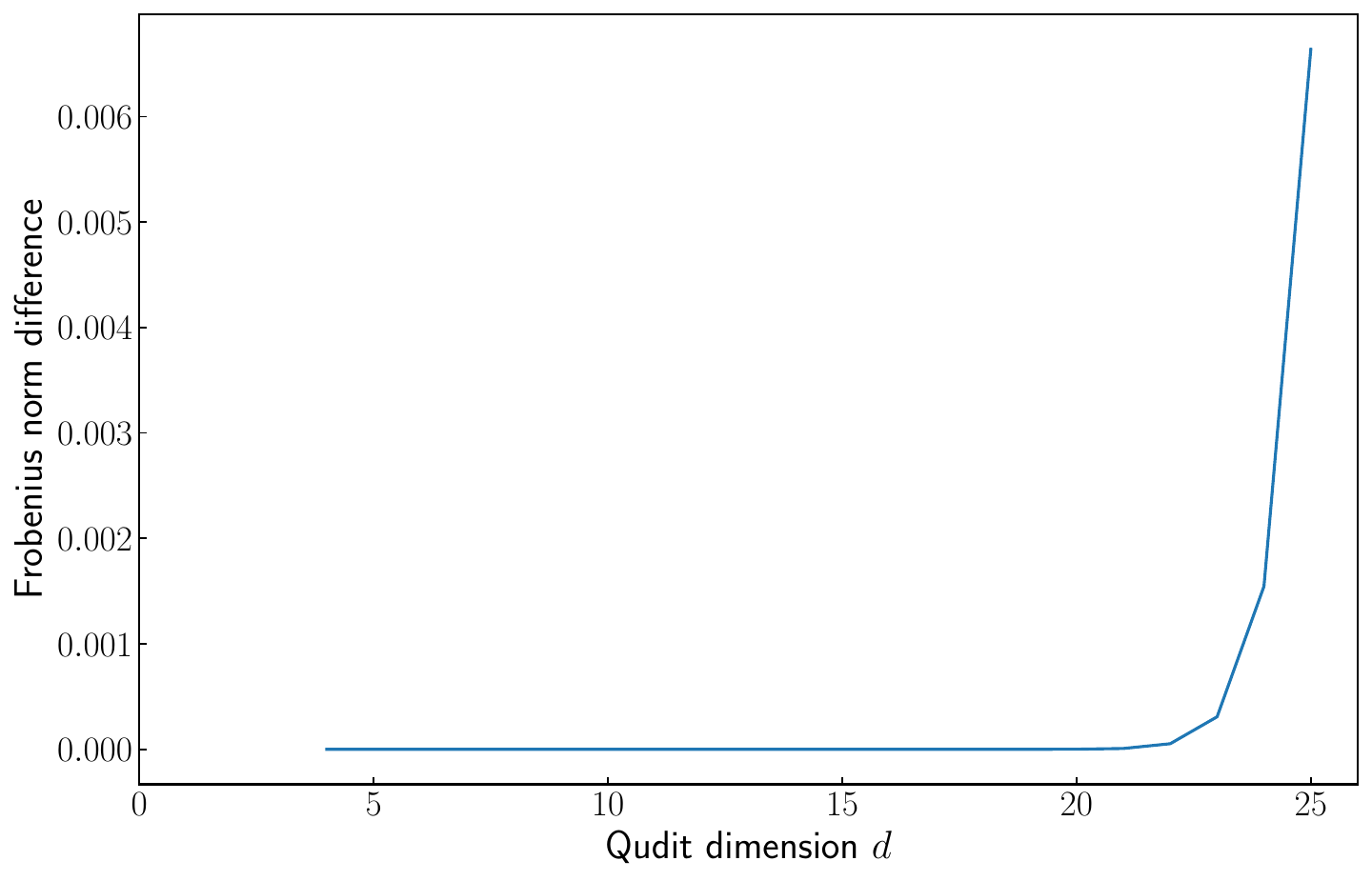}
\caption{Average Frobenius norm difference between the true and truncated versions of a 100 displacement operators for each value of qudit dimension, with the displacement parameter being sampled uniformly randomly in $(-5.0, 5.0)$.}
\label{fig:diff_true_truncated_disp}
\end{figure}

\begin{figure}[!t]
\centering
\includegraphics[width=0.5\linewidth]{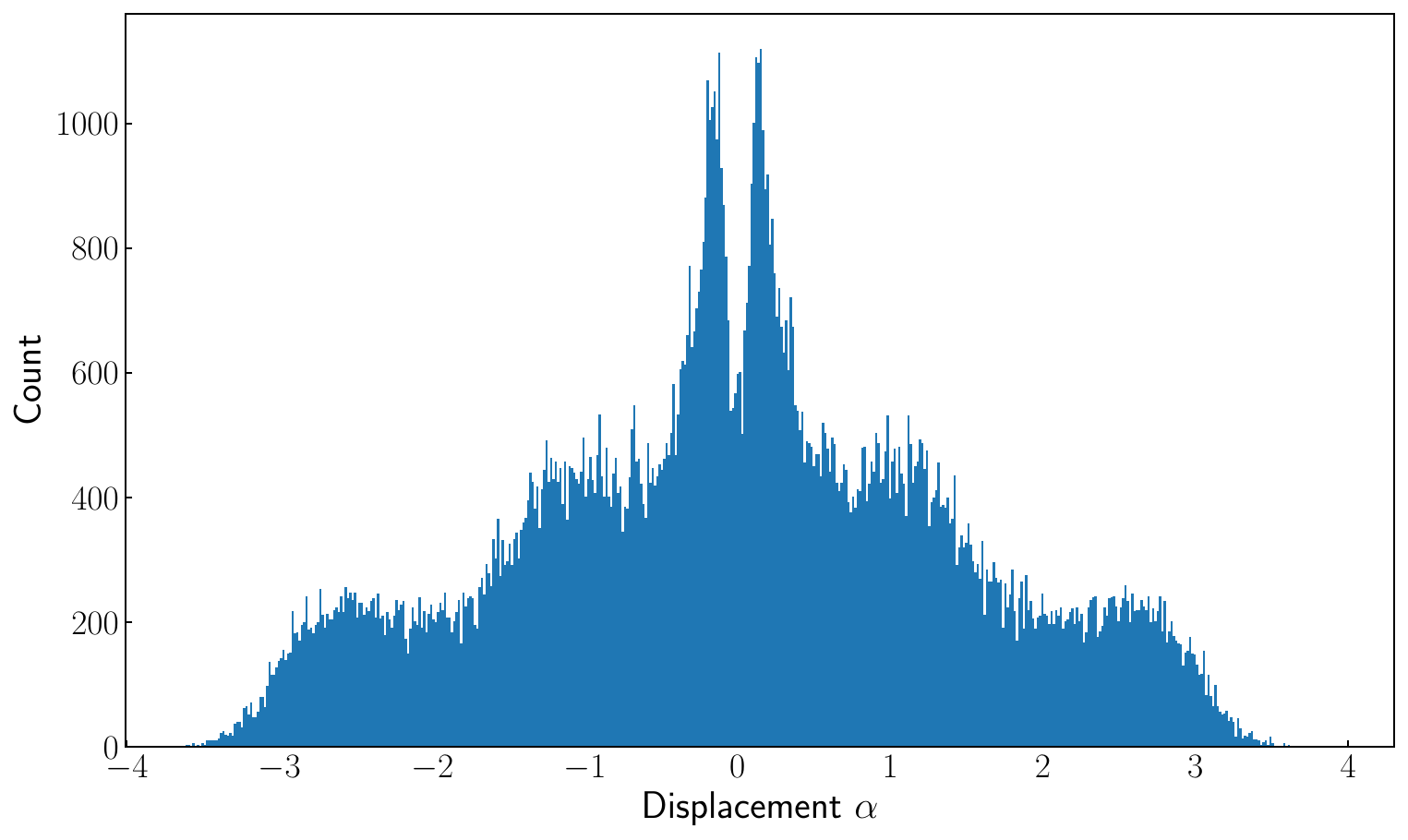}
\caption{Distribution of displacement gate parameters appearing in the ansatz, Eq.~\ref{eqn:snapdisplacementansatz} (with $k=2$), arising from our parameter solver, aggregated over simulations with a variety of qubit T${}_1$'s, qudit dimensions and 1000's of unitaries each.}
\label{fig:displacement_distribution}
\end{figure}

The displacement gate, defined in terms of the creation $(a^\dagger)$ and annihilation $(a)$ operators
\begin{equation}
\hat{D}(\alpha) = e^{\alpha \hat{a}^{\dagger} - \alpha^{\star}\hat{a}},
\end{equation}
creates coherent states out of the vacuum
\begin{equation}
\ket{\alpha} = \hat{D}(\alpha) \ket{0},
\end{equation}
where the coherent states are defined as eigenstates of the annihilation operator $\hat{a} \ket{\alpha} = \alpha \ket{\alpha}$, and have an expansion in terms of the Fock basis number states as
\begin{equation}
\ket{\alpha} = \text{exp}\left( -\frac{1}{2} \vert \alpha \vert^2 \right) \sum_{n=0}^{\infty} \frac{\alpha^{n}}{n!} \ket{n} .
\end{equation}
In the Fock basis, the displacement gate has matrix elements given by \cite{cahill1969ordered}
\begin{equation}
\bra{m} \hat{D}(\alpha) \ket{n} = \sqrt{\frac{n!}{m!}} \alpha^{m-n} e^{-\vert \alpha\vert^2/2} L_{n}^{m-n} \left(\vert \alpha \vert^2 \right),
\end{equation}
where
\begin{equation}
L_{n}^{k}(x) = \frac{1}{n!} \sum_{j=0}^{n} \frac{n!}{j!} \binom{k+n}{n-j} (-x)^{j},
\end{equation}
is the associated Laguerre polynomial, and where $\binom{n}{k}$ is a binomial coefficient.

In practice, we must truncate the Fock basis up to some finite number $N_{\rm cavity}$, which needs to be large enough such that in the computational subspace $d \ll N_{\rm cavity}$ of the qudit, the matrix elements above agree to some negligible error. The true displacement operator acts in a space where $N_{\rm cavity} \rightarrow \infty$, so that even in the absence of any physical errors, or even compilation errors, the optimized parameters we find may not correspond to the ideal parameters that would reproduce the desired displacement operator which operators in a space of unbounded cavity dimension. This residual error is attributable purely to the truncation of this cavity dimension, and so we ensure that it is negligibly small.

In our simulations, we chose to truncate the cavity dimension to $N_{\rm cavity} = 60$, and limited our simulations to qudit dimensions of up to 25. Figure~\ref{fig:diff_true_truncated_disp} shows the average norm difference between the truncated and true displacement operators, taken across 100 displacement operators for each value of the qudit dimension, with the displacement parameters draw uniformly randomly between -5 and 5. Here, we take the Frobenius norm, which for some operator $A$ is defined as
\begin{equation}
\Vert A \Vert_{2} = \left[ \sum_{i,j} |a_{i,j}|^{2} \right]^{1/2}.
\end{equation}

The choice of the range for uniform sampling of the displacement parameters is guided by the numerical range of the optimized displacement operators. Note that in our ansatz, the displacement parameters can be taken to be real. The distribution of the optimized parameters is plotted in Fig.~\ref{fig:displacement_distribution} With the exception of a few outliers, which we ignore, the overwhelming majority of the optimized displacement parameters fall within the range $(-3.98, 3.86)$. We round this interval to $(-5, 5)$ in our estimate of the errors arising due to truncation above.

\end{document}